\documentstyle[11pt,epsf,dina4p]{article}
\begin{document}
%
%---- filenames for plots
%
\newcommand{\figI}   	{mt_tb0}          	% m_top vs. tanb
\newcommand{\figII}  	{m1_m2}           	% m_1/m_2 vs. Q
\newcommand{\figIII} 	{chi2_gyebdl}     	% chi^2 gyebdl
\newcommand{\figIV}   	{spectrum}        	% spectrum(Q)
\newcommand{\figV}   	{bsg}             	% b-->sgamma
\newcommand{\figVI}  	{relic}           	% relic density
\newcommand{\figVII} 	{mchar1}          	% m_charg
\newcommand{\figVIII}  	{chi2_gyel}       	% chi^2 gyel
\newcommand{\figIX}  	{dmb}             	% Delta_mb
\newcommand{\figX}   	{mt_tb1}          	% m_top vs. tanb +/-Dmb, +/-bsg
\newcommand{\figXa} 	{running_At}       	% running of At
\newcommand{\figXI} 	{mglui}           	% Gluino mass m_A
\newcommand{\figXII} 	{mu_mglu}         	% mu/m_gluino
\newcommand{\figXIII}	{atop_mglu}       	% atop/m_gluino
\newcommand{\figXIV} 	{abot_mglu}    	 	% abot/m_gluino
\newcommand{\figXV} 	{atau_mglu}       	% atau/m_gluino
\newcommand{\figXVI}  	{mu}              	% Mu
\newcommand{\figXVII} 	{mh}              	% light CP even Higgs mass m_h
\newcommand{\figXVIII}	{ma}              	% CP odd Higgs mass m_A
\newcommand{\figXIXa}	{m1}	                % m1
\newcommand{\figXIXb}	{m2}	                % m2
\newcommand{\figXIXc}	{S1_m1}	                % sigma1/m1
\newcommand{\figXIXd}	{S2_m2}	                % sigma2/m2
\newcommand{\figXIXe}   {dmz}                   % Delta_MZ vs m0,m1/2
\newcommand{\figXX}  	{43a_hz0}       	% higgs mass,x-section
\newcommand{\figXXI} 	{49a_hz0-300}       	% higgs mass,x-section
\newcommand{\figXXII} 	{x_hz0_low_high}        % higgs mass,x-section
\newcommand{\figXXIII} 	{49a_chichi-300}	% charg,neutral x-sections

\newcommand{\unify}  {plot1}           % Unification of couplings
\newcommand{\runma}  {plot5}           % running of SUSY masses
\newcommand{\proton} {proton}          % 90% excluded region of proton decay
\newcommand{\mbvsmt} {plot3}           % mb vs. mtop for different tanb
\newcommand{\mhvsmt} {plot7}           % mh vs. mtop for diff. tanb,m0,m1/2
\newcommand{\chisq}  {chi2}            % chi^2 vs m0,m1/2
\newcommand{\chisqp} {chi2prot}        % chi^2 vs m0,m1/2 incl. proton decay
\newcommand{\mufit}  {mu}              % mu vs m0,m1/2 tanb=2
\newcommand{\mucormh}{plot4}           % chi^2 vs mu,m1/2 tanb=2
%---- miscellaneous abrreviations
%
\newcommand{\bq}{\begin{equation}}
\newcommand{\eq}{\end{equation}}
\newcommand{\beq}  {\begin{eqnarray}}
\newcommand{\eeq}  {\end{eqnarray}}
\newcommand{\rG}   {{\rm GUT}}
\newcommand{\MG}   {{\ifmmode M_\rG         \else $M_\rG$          \fi}}
\newcommand{\mb}   {{\ifmmode m_{b}         \else $m_{b}$          \fi}}
\newcommand{\mt}   {{\ifmmode m_{t}         \else $m_{t}$          \fi}}
\newcommand{\agut} {{\ifmmode \alpha_\rG    \else $\alpha_\rG$     \fi}}
\newcommand{\mgut} {{\ifmmode M_\rG         \else $M_\rG$          \fi}}
\newcommand{\mze}  {{\ifmmode m_0           \else $m_0$            \fi}}
\newcommand{\mha}  {{\ifmmode m_{1/2}       \else $m_{1/2}$        \fi}}
\newcommand{\tb}   {{\ifmmode \tan\beta     \else $\tan\beta$      \fi}}
\newcommand{\mz}   {{\ifmmode M_{Z}         \else $M_{Z}$          \fi}}
\newcommand{\ai}   {{\ifmmode \alpha_i      \else $\alpha_i$       \fi}}
\newcommand{\aii}  {{\ifmmode \alpha_i^{-1} \else $\alpha_i^{-1}$  \fi}}
\newcommand{\DRbar}{{\ifmmode \overline{DR} \else $ \overline{DR}$ \fi}}
\newcommand{\msusy}{{\ifmmode M_{SUSY}      \else $M_{SUSY}$       \fi}}
\newcommand{\as}   {{\ifmmode \alpha_s      \else $\alpha_s$       \fi}}
\newcommand{\asmz} {{\ifmmode \alpha_s(M_Z) \else $\alpha_s(M_Z)$  \fi}}
\newcommand{\tal}  {{\ifmmode \tilde{\alpha} \else $\tilde{\alpha}$ \fi}}
\newcommand{\rb}[1]{\raisebox{1.5ex}[-1.5ex]{#1}}
\newcommand {\tabs}[1]{\multicolumn{1}{c}{\mbox{\hspace{#1}}}}
\newcommand{\sws}  {{\ifmmode \;\sin^2\theta_W
                     \else    $\;\sin^{2}\theta_{W}$               \fi}}
\newcommand{\cws}  {{\ifmmode \;\cos^2\theta_W
                     \else    $\;\cos^{2}\theta_{W}$               \fi}}
\newcommand{\sw}   {{\ifmmode\;\sin\theta_W\else $\sin\theta_{W}$  \fi}}
\newcommand{\cw}   {{\ifmmode\;\cos\theta_W\else $\;\cos\theta_{W}$\fi}}
\newcommand{\tw}   {{\ifmmode\;\tan\theta_W\else $\;\tan\theta_{W}$\fi}}
\newcommand{\bsg}  {{\ifmmode \b\rightarrow s\gamma
\else $b\rightarrow s\gamma$ \fi}}
\newcommand{\Bbsg}  {{\ifmmode BR(\b\rightarrow s\gamma)
\else $BR(b\rightarrow s\gamma)$ \fi}}
\newcommand{\smas}[2]{\tilde{m}^#2_{#1}}
\newcommand{\nn}   {\nonumber \\}
\renewcommand{\floatpagefraction}{0.005}

%%%%%%%%%%%%%%%%%%%%%%%
% End of Declarations J.M %
%%%%%%%%%%%%%%%%%%%%%%%
\newcommand{\Zto}   {\mbox{$\mathrm Z \to$}}

\def\NPB#1#2#3{{\it Nucl.~Phys.} {\bf{B#1}} (19#2) #3}
\def\PLB#1#2#3{{\it Phys.~Lett.} {\bf{B#1}} (19#2) #3}
\def\PRD#1#2#3{{\it Phys.~Rev.} {\bf{D#1}} (19#2) #3}
\def\PRL#1#2#3{{\it Phys.~Rev.~Lett.} {\bf{#1}} (19#2) #3}
\def\ZPC#1#2#3{{\it Z.~Phys.} {\bf C#1} (19#2) #3}
\def\PTP#1#2#3{{\it Prog.~Theor.~Phys.} {\bf#1}  (19#2) #3}
\def\MPL#1#2#3{{\it Mod.~Phys.~Lett.} {\bf#1} (19#2) #3}
\def\PR#1#2#3{{\it Phys.~Rep.} {\bf#1} (19#2) #3}
\def\RMP#1#2#3{{\it Rev.~Mod.~Phys.} {\bf#1} (19#2) #3}
\def\HPA#1#2#3{{\it Helv.~Phys.~Acta} {\bf#1} (19#2) #3}
\def\NIMA#1#2#3{{\it Nucl.~Instr.~and~Meth.} {\bf#1} (19#2) #3}

\begin{titlepage}
\begin{flushright}
\vspace*{-2.2cm}
\noindent
EPS-HEP 95 Ref. eps0631         \hfill
           IEKP-KA/95-07    \\
%Submitted to Pa 06,09,13,15~Pl 12,21 \hfill June 30th, 1995  \\%
 \hfill hep-ph/9507291     \\
%\hspace*{2.35cm}Pl\hspace{2pt} 12, 21
%%\hfill
 %      June 30th, 1995        \\
\end{flushright}
\vspace{0.5cm}
\begin{center} {\Large\bf Combined Fit of Low Energy Constraints to   \\
          Minimal Supersymmetry and Discovery Potential at LEP II\\}
\vspace{0.3cm}
{\bf W.~de Boer\footnote{Email: DEBOERW@CERNVM},
 G.~Burkart\footnote{E-mail: gerd@ekpux8.physik.uni-karlsruhe.de},
 R.~Ehret\footnote{E-mail: ehret@ekpux7.physik.uni-karlsruhe.de},
 W.~Oberschulte-Beckmann\footnote{E-mail: wulf@ekpux5.physik.uni-karlsruhe.de}
\\
}
{\it Inst.\ f\"ur Experimentelle Kernphysik, Univ.\ of Karlsruhe   \\}
{\it Postfach 6980, D-76128 Karlsruhe 1, FRG  \\} and \\
%{\bf V. Bednyakov, A.V.~Gladyshev, D.I.
%Kazakov\footnote{E-mail:
% kazakovd@thsun1.jinr.dubna.su}, S.G. Kovalenko\footnote{E-mail:
%kovalen@lnpnw1.jinr.dubna.su}\\}
{\bf V. Bednyakov,  S.G. Kovalenko\footnote{E-mail:
kovalen@lnpnw1.jinr.dubna.su}\\}
{\it Bogoliubov Lab. of Theor. Physics,
     Joint Inst. for Nucl. Research, \\}
{\it 141 980 Dubna, Moscow Region, RUSSIA \\}
\end{center}

\vspace{0.3cm}

\begin{center}
{\bf Abstract}
\end{center}
\vspace{0.3cm}
\parindent0.0pt
\small
 Within the Constrained Minimal Supersymmetric Standard Model (CMSSM)
it is possible to predict the low energy gauge couplings and   masses
of the 3.~generation particles from a few parameters at
the GUT scale. In addition the MSSM predicts  electroweak symmetry
breaking due to large radiative corrections from Yukawa couplings, thus
relating the $Z^0$ boson mass to the top quark mass.

{}From a $\chi^2$~analysis, in which these  constraints
are considered simultaneously,  one can calculate the
probability for each point in the MSGUT parameter space.
The recently measured top quark mass prefers two solutions
for the mixing angle in the Higgs sector:  $\tan\beta$ in
the range between 1 and 3 or alternatively
$\tan\beta\approx 15-50$.
For both cases we find  a unique $\chi^2$ minimum in the
 parameter space. From the  corresponding
 most probable parameters at the GUT scale,
the masses of all predicted particles  can be calculated
at low energies using the RGE, albeit with rather large
errors due to the logarithmic nature of the running of
the masses and coupling constants.
Our fits  include full second order corrections
for the gauge and Yukawa couplings, low energy threshold effects,
 contributions of all (s)particles to the Higgs potential
 and corrections to $m_b$ from gluinos and
higgsinos, which exclude  (in our notation) positive
values of the mixing parameter in the Higgs potential
$\mu$ for the large $\tan\beta$ region.

Further constraints can be derived  from~ the
branching ratio for the radiative (penguin) decay of the
b-quark into $s\gamma$ %, the proton decay limits
and   the lower limit on the lifetime of the universe,
which requires the     dark matter density due to the Lightest
Supersymmetric Particle (LSP) not to overclose the universe.

For the low $\tan\beta$ solution these additional
constraints can be fulfilled simultaneously for quite
a large region of the parameter space. In contrast,
for the high $\tan\beta$ solution the correct value for the
$ b  \rightarrow s\gamma$ rate is obtained
  only for small values of the gaugino scale
%In the latter case  %the proton decay is too fast,
and  electroweak symmetry breaking is difficult,
unless one assumes the minimal SU(5) to be a subgroup
of a larger symmetry group, which is broken
between the Planck scale and the unification scale.
In this case small splittings in the Yukawa couplings
are expected  at the unification scale  and electroweak
symmetry breaking is easily obtained,
provided the Yukawa coupling for the top quark
is slightly above the one for the bottom quark, as
expected e.g. if the larger symmetry group would
be SO(10).
% and usually very little dark matter is predicted.
%Clearly, these additional constraints prefer the low
%$\tan\beta$ solution, but one should keep in mind that
%the higher orders for the $ b  \rightarrow s\gamma$ rate,
%which have not yet been calculated, might be large,
%%the proton decay is model  dependent,
%and the dark matter prediction is very sensitive to the
%mixing in the neutralino sector. This mixing
%is large for the high $\tan\beta$ solution due to the constraints
%from  $ b  \rightarrow s\gamma$  and electroweak
%symmetry breaking.

For   particles, which are most likely to have masses
in the LEP II energy range, the cross sections are given
for the various energy scenarios at LEP II.
The highest LEP II energies (205 GeV) are just high enough to
cover a large region of the preferred parameter space,
both for the low and high $\tb$ solutions.
 For low $\tb$   the production of the lightest
 Higgs boson, which is expected to have a mass below 115 GeV,
is the most promising channel, while for  large $\tb$
the production of Higgses, charginos and/or neutralinos
covers the preferred parameter  space.

%\vspace{0.5cm}
%\begin{center}
%(Submitted to
% Physics Letters B.)
%\end{center}
\end{titlepage}
\section{Introduction}
Grand Unified Theories (GUT's) in which the electroweak and
strong forces are unified at a scale \MG of the order $10^{16}$
GeV are strongly constrained by low energy data, if one imposes
unification of gauge- and Yukawa couplings as well as
 electroweak symmetrybreaking.
The Minimal Supersymmetric Standard Model (MSSM) \cite{su5susy}
 has become the
leading candidate for a GUT after the precisely measured coupling
constants at LEP excluded   unification in the Standard Model
\cite{ekn2,abf,lalu}.
In the MSSM the quadratic divergences in the higher order radiative
corrections largely cancel, so one can calculate the corrections
reliably even over many orders of magnitude.
The large hierarchy between the electroweak scale
and the unification scale as well as the
different strengths of the forces at low
energy are naturally explained by the radiative corrections\cite{rev}.
Low energy data on masses and couplings provide strong constraints
on the MSSM parameter space, as discussed recently by many groups
{}~\cite{rrb92,cpw,bbo,op,chan,langpol,zic,wdb,bek,nanopo,ram1,roskane,roskane1,cawa,copw}.

In this paper we perform a combined statistical analysis of the low energy
constraints, namely the three gauge coupling constants measured at LEP,
the quark- and leptonmasses of the third generation, the lower limit
on the as yet unobserved supersymmetric particles, the $Z^0$-boson mass,
 the radiative decay \bsg observed by CLEO~\cite{cleo94}, and the lower limit
on the lifetime of the universe, which requires the dark matter
density from  the Lightest Supersymmetric Particle (LSP)
not to  overclose the universe.
No restriction is made on \tb, the ratio of vacuum expectation
values of the neutral components of the Higgsfields. Therefore
both the low \tb solution, expected in the SU(5), and the high \tb
solution, expected in SO(10), are considered.

The theoretically more questionable constraint  from proton decay in the MSSM
 \cite{arn,langac},
which involve the unknown Higgs sector at the GUT scale,
was    considered in a similar analysis before\cite{bek}.
At large $\tb$ values one needs a different multiplet structure\cite{flipsu5}
or a larger Higgs sector\cite{finite}.

Assuming soft symmetry breaking at the \rG-scale, all SUSY masses
can be expressed in terms of 5 parameters and the masses at low
energy are then determined by the well known Renormalization
Group Equations (RGE). The experimental constraints are sufficient
to determine these parameters, albeit with large uncertainties.
{}From the statistical analysis we obtain the probability for every
point in the SUSY parameter space, which allows us to calculate
the cross sections for the expected new physics of the MSSM at LEP II.
 These cross sections will be given as function of the common
scalar and gaugino masses at the \rG-scale, denoted by \mze, \mha;
 for each choice of \mze, \mha, the other parameters were determined
from the constraint fit.
\section{The Model}
\subsection{The Lagrangian}
The minimal supersymmetric extension of the Standard Model is described
by the Lagrangian containing the SUSY-symmetric part together with
SUSY breaking terms originating from supergravity~\cite{susy}.
The breaking terms of the Lagrangian are given by:
\begin{eqnarray} {\cal L}_{Breaking} & = &
-m_0^2\sum_{i}^{}|\varphi_i|^2-m_{1/2}\sum_{j}^{}\lambda_j
 \lambda_j \label{2} \\ & - &
   Am_0\left[h_u^{ab}Q_aU^c_bH_2+h_d^{ab}Q_aD^c_bH_1+
 h_e^{ab}L_aE^c_bH_1\right] - Bm_0\left[\mu H_1H_2\right].
 \nonumber \end{eqnarray}
\noindent
The Lagrangian given above assumes that
  squarks and sleptons  have a common mass $m_0$
and the gauginos a common mass $m_{1/2}$ at the GUT scale.
The   SUSY Lagrangian contains the following free parameters:
\begin{itemize}
\item 3 gauge couplings $\alpha_i$,
\item the Yukawa couplings $h_i^{ab}$, where $i$ is the flavour index
and $ab$ are generation indices. Since   the masses of the third
generation are much larger than masses of the first two ones, we consider
only the Yukawa coupling of the third geneeration and drop the indices $a,b$.
 \item the Higgs field mixing parameter  $\mu $.
\end{itemize}
\noindent
They are supplemented by the soft breaking ones:
\begin{itemize}
\item $m_0, \ m_{1/2}, \ A, \ B$,
where A and B are the coupling constants
for the Higgs fields.
\end{itemize}
With these parameters the comlete mass spectrum of the
SUSY particles is determined.
\subsection{The SUSY Mass Spectrum \label{rge}}
All couplings and masses become scale dependent due to
radiative corrections. This running is described by
the renormalization group equations (RGE)~\cite{bbo}:\\

{\bf 3 Couplings: ($i=1,2,3$)} \\
\begin{eqnarray}
   \frac{d\tal_{i}}{dt} & = &
            -b_i \tal_i^2 -\tal_i^2\left( \sum_{j=1}^3
            b_{ij}\tal_j- \sum_{U,D,L}a_{ij} Y_j\right)
                                             \label{coupl}   \\
   \frac{dY_U}{dt}    & = & Y_U
           \sum_{i=1}^6 \left( c^t_i \tal_i -
           \sum_{j\geq i}^6 c^t_{ij} \tal_i \tal_j \right)
                                                \label{ytop} \\
   \frac{dY_D}{dt}    & = & Y_D
           \sum_{i=1}^6 \left( c^b_i \tal_i -
           \sum_{j\geq i}^6 c^b_{ij} \tal_i \tal_j \right)
                                                \label{ybot} \\
   \frac{dY_L}{dt} & = & Y_L
           \sum_{i=1}^6 \left( c^\tau_i \tal_i -
           \sum_{j\geq i}^6 c^\tau_{ij} \tal_i \tal_j \right)
\end{eqnarray}
{\bf 3 Gauginos: ($i=1,2,3$)} \\
\begin{eqnarray}
    \frac{dM_{i}}{dt} & = & -b_i\tilde{\alpha}_i^2M_i        \label{RGEM}
\end{eqnarray}
{\bf Masses of the 1st. and 2nd Generation ($i=1,2$):} \\
\begin{eqnarray}
\frac{d\tilde{m}^2_{L_i}}{dt} & = &
    3(\tilde{\alpha}_2M^2_2 + \frac{1}{5}\tilde{\alpha}_1M^2_1) \\
\frac{d\tilde{m}^2_{E_i}}{dt} & = & (
     \frac{12}{5}\tilde{\alpha}_1M^2_1) \\
\frac{d\tilde{m}^2_{Q_i}}{dt} & = & (\frac{16}{3}\tilde{\alpha}_3M^2_3
    + 3\tilde{\alpha}_2M^2_2 + \frac{1}{15}\tilde{\alpha}_1M^2_1)\\
\frac{d\tilde{m}^2_{U_i}}{dt} & = & (\frac{16}{3}\tilde{\alpha}_3M^2_3
    +\frac{16}{15}\tilde{\alpha}_1M^2_1)  \\
\frac{d\tilde{m}^2_{D_i}}{dt} & = &
    (\frac{16}{3}\tilde{\alpha}_3M^2_3+
    \frac{4}{15}\tilde{\alpha}_1M^2_1)
\end{eqnarray}
{\bf Masses of the 3th Generation ($i=3$):} \\
\begin{eqnarray}
\frac{d\tilde{m}^2_{L_i}}{dt} & = &
    3(\tilde{\alpha}_2M^2_2 + \frac{1}{5}\tilde{\alpha}_1M^2_1)
   -Y_L(\tilde{m}^2_L+\tilde{m}^2_E+m^2_{H_D}+A^2_Lm_0^2)
           \label{RGEmL}\\
\frac{d\tilde{m}^2_{E_i}}{dt} & = & (
     \frac{12}{5}\tilde{\alpha}_1M^2_1)
    -2Y_L(\tilde{m}^2_L+\tilde{m}^2_E
    +m^2_{H_D}+A^2_Lm_0^2) \label{RGEmE}\\
\frac{d\tilde{m}^2_{Q_i}}{dt} & = & (\frac{16}{3}\tilde{\alpha}_3M^2_3
    + 3\tilde{\alpha}_2M^2_2 + \frac{1}{15}\tilde{\alpha}_1M^2_1)-
      \label{RGEmQ} \\
 &&  \left[
     Y_U(\tilde{m}^2_Q+\tilde{m}^2_U+m^2_{H_U}+A^2_Um_0^2)
    +Y_D(\tilde{m}^2_Q+\tilde{m}^2_D+m^2_{H_D}+A^2_Dm_0^2)
     \right],                                                     \nn
\frac{d\tilde{m}^2_{U_i}}{dt} & = & (\frac{16}{3}\tilde{\alpha}_3M^2_3
    +\frac{16}{15}\tilde{\alpha}_1M^2_1)
    -2Y_U(\tilde{m}^2_Q+\tilde{m}^2_U+
                            m^2_{H_U}+A^2_Um_0^2), \label{RGEmU}\\
\frac{d\tilde{m}^2_{D_i}}{dt} & = &
    (\frac{16}{3}\tilde{\alpha}_3M^2_3+
    \frac{4}{15}\tilde{\alpha}_1M^2_1)
    -2Y_D(\tilde{m}^2_Q+\tilde{m}^2_D+m^2_{H_D}+A^2_Dm_0^2)
                  \label{RGEmD}
\end{eqnarray}
{\bf Higgs potential parameters:}\\
\begin{eqnarray}
\frac{d\mu^2}{dt}&=&\mu^2\left[3(\tilde{\alpha}_2+
\frac{1}{5}\tilde{\alpha}_1)-(3Y_U+3Y_D+Y_L)\right] \\
\frac{dm^2_1}{dt} & = &
3(\tilde{\alpha}_2M^2_2 +\frac{1}{5}\tilde{\alpha}_1M^2_1)+
3(\tilde{\alpha}_2 +\frac{1}{5}\tilde{\alpha}_1)\mu^2-(3Y_U+3Y_D+Y_L)
\mu^2 \\
& & -3Y_D(\tilde{m}^2_Q+\tilde{m}^2_D+m^2_1-\mu^2+A^2_Dm_0^2)-
Y_E(\tilde{m}^2_L+\tilde{m}^2_E+m^2_1-\mu^2+A^2_Lm_0^2) \\
\frac{dm^2_2}{dt} & = &
3(\tilde{\alpha}_2M^2_2 +\frac{1}{5}\tilde{\alpha}_1M^2_1)+
3(\tilde{\alpha}_2
+\frac{1}{5}\tilde{\alpha}_1)\mu^2-(3Y_U+3Y_D+Y_E)\mu^2 \\
& & -3Y_U(\tilde{m}^2_Q+\tilde{m}^2_U+m^2_2-\mu^2+A^2_Um_0^2) \\
%
%\frac{dm^2_3}{dt} & = & \frac{3}{2}(
% \tilde{\alpha}_2+\frac{1}{5}\tilde{\alpha}_1-Y_t)m^2_3+3\mu m_0Y_tA_t-
%3\mu(\tilde{\alpha}_2M_2 +\frac{1}{5}\tilde{\alpha}_1M_1)\\
%                      %! Correction to dmitri2
\end{eqnarray}
{\bf Trilinear couplings:} \\
\begin{eqnarray}
\frac{dA_U}{dt} & = &
  -\left(\frac{16}{3}\tilde{\alpha}_3\frac{M_3}{m_0} +
  3\tilde{\alpha}_2\frac{M_2}{m_0} +
  \frac{13}{15}\tilde{\alpha}_1\frac{M_1}{m_0}\right)-6Y_UA_U-Y_DA_D\\
\frac{dA_D}{dt} & = &
  -\left(\frac{16}{3}\tilde{\alpha}_3\frac{M_3}{m_0}
  + 3\tilde{\alpha}_2\frac{M_2}{m_0} +
  \frac{7}{15}\tilde{\alpha}_1\frac{M_1}{m_0}\right)-
  6Y_DA_D-Y_UA_U-Y_LA_L\\
\frac{dA_L}{dt} & = & -\left(
 3\tilde{\alpha}_2\frac{M_2}{m_0} +
\frac{9}{5}\tilde{\alpha}_1\frac{M_1}{m_0}\right)-3Y_DA_D-4Y_LA_L\\
%
%
%\frac{dB}{dt} & =
%& -3\left(\tilde{\alpha}_2\frac{M_2}{m_0} +
%\frac{1}{5}\tilde{\alpha}_1\frac{M_1}{m_0}\right)-3Y_UA_U -3Y_DA_D-Y_LA_L
\end{eqnarray}
Here $\tilde{m}_U,\tilde{m}_D$ and $\tilde{m}_E$ refer to the masses of the
superpartners
of the quark and lepton singlets, while $\tilde{m}_Q$ and $\tilde{m}_L$
refer to the
masses of the weak isospin doublet superpartners;  $m_1, m_2, m_3$ and
$\mu$ are the mass parameters of the Higgs
potential (see next section),
while $A_U,A_D,A_L$ and $B$ are the couplings in
${\cal L}_{Breaking}$ as defined before;  $M_i$ are the gaugino
masses before any mixing and the following notation is used:
$$ \tal_i=\frac{\alpha_i}{4\pi},
\ t=\log(\frac{M_X^2}{Q^2}),Y_j=\frac{h_j^2}{(4\pi)^2}, $$
where $i=1,2,3$ and $j=U,D,L$. Only the Yukawa couplings of the third
generation are considered, so $Y_U, Y_D, Y_L$ refer  to $Y_t, Y_b, Y_\tau$ and
the  couplings $h_j$ are related to the masses by
\beq\label{yukawastbtau}
 m_{t}&=&h_t(m_t)v\sin\beta \nn
 m_b & =& h_b(m_b)v\cos\beta \nn
 m_\tau&=& h_\tau(m_\tau)v\cos\beta
\eeq
Here $m_j$ are the running masses.
The boundary conditions at $Q^2=M_{\rm GUT}^2$ or at $t=0$ are:
$$\tilde{m}^2_Q=\tilde{m}^2_U=\tilde{m}^2_D=\tilde{m}^2_L=\tilde{m}^2_E=
m_0^2;$$
$$\mu^2 = \mu_0^2;\ \ \  m_1^2=m^2_2=\mu_0^2+m_0^2; \ \ \ $$
$$M_i=m_{1/2};\ \ \ \tal_i(0)=\tal_{\rm GUT},\ \ \ i=1,2,3$$
With given values for $m_0,m_{1/2},\mu,Y_t,Y_b,Y_\tau,\tan\beta$,
and A and correspondingly known boundary conditions at the GUT scale,
the differential equations can be solved numerically thus linking the
values at the GUT and electroweak scales. The non-negligible
Yukawa couplings cause   a mixing between the
electroweak eigenstates and the mass
eigenstates of the third generation particles.
The mixing matrices for the $\smas{t}{2},\smas{b}{2}$
and $\smas{\tau}{2}$
 are:
\begin{equation} \label{stopmat}
\left(\begin{array}{cc}
\tilde{m}_Q^2+m_t^2+\frac{1}{6}(4M_W^2-M_Z^2)\cos 2\beta &
m_t(A_tm_0-\mu\cot \beta ) \\
m_t(A_tm_0-\mu\cot \beta ) &
\tilde{m}_U^2+m_t^2-\frac{2}{3}(M_W^2-M_Z^2)\cos 2\beta
\end{array}  \right)       \nonumber
\end{equation}
\begin{equation} \label{botmat}
\left(\begin{array}{cc}
\tilde{m}_Q^2+m_b^2-\frac{1}{6}(2M_W^2+M_Z^2)\cos 2\beta &
m_b(A_bm_0-\mu\tan \beta ) \\
m_b(A_bm_0-\mu\tan \beta ) &
\tilde{m}_D^2+m_b^2+\frac{1}{3}(M_W^2-M_Z^2)\cos 2\beta
\end{array}  \right)               \nonumber
\end{equation}
\begin{equation} \label{staumat}
\left(\begin{array}{cc}
\tilde{m}_L^2+m_{\tau}^2-\frac{1}{2}(2M_W^2-M_Z^2)\cos 2\beta &
m_{\tau}(A_{\tau}m_0-\mu\tan \beta ) \\
m_{\tau}(A_{\tau}m_0-\mu\tan \beta ) &
\tilde{m}_E^2+m_{\tau}^2+(M_W^2-M_Z^2)\cos 2\beta
\end{array}  \right)               \nonumber
\end{equation}
and the  mass eigenstates are
 the eigenvalues of these mass matrices.
The mass matrix for the  neutralinos can be written in our notation
as:
\begin{equation}
  {\cal M}^{0}=\left(
  \begin{array}{cccc}
    M_1 & 0   &-M_Z\cos\beta \sin_W & M_Z\sin\beta \sin_W  \\
    0   & M_2 & M_Z\cos\beta \cos_W &-M_Z\sin\beta \cos_W  \\
   -M_Z\cos\beta \sin_W & M_Z\cos\beta \cos_W & 0   & -\mu \\
    M_Z\sin\beta \sin_W &-M_Z\sin\beta \cos_W &-\mu & 0
  \end{array} \right) \label{neutmix} \\
\end{equation}
The physical neutralino masses  $M_{\tilde{\chi}_i^0}$
are obtained as eigenvalues of this matrix after diagonalization.
The mass matrix for the charginos is:
\begin{equation}
  {\cal M}^{\pm}=\left(
  \begin{array}{cc}
     M_2                  & \sqrt{2}M_W\sin\beta \\
     \sqrt{2}M_W\cos\beta & \mu
  \end{array} \right) \label{charmix}
\end{equation}
This matrix has two eigenvalues
corresponding to the masses of the two charginos~ $\tilde{\chi}_{1,2}^{\pm}$:
$$
  M^2_{1,2}=\frac{1}{2}\left[M^2_2+\mu^2+2M^2_W \mp
  \sqrt{(M^2_2-\mu^2)^2+4M^4_W\cos^22\beta +4M^2_W(M^2_2+\mu^2+2M_2\mu
  \sin 2\beta )}\right]
$$

\subsection{Radiative Corrections to the Higgs potential}
\label{cor_higgs_pot}
The Higgs potential $V$ including the one-loop corrections $ \quad\Delta V$
can be written as:
\begin{eqnarray}
  V(H_1^0,H_2^0) &=& m^2_1|H_1^0|^2+m^2_2|H_2^0|^2-
                     m^2_3(H_1^0H_2^0+h.c.)+
\frac{g^2+g^{'2}}{8}(|H_1^0|^2-|H_2^0|^2)^2 + \quad\Delta V\nonumber  \\
{\rm with }\quad\Delta V&=&\frac{1}{64\pi^2}
\sum_i(-1)^{2J_i}(2J_i+1)C_im_i^4
\left[\ln\frac{m_i^2}{Q^2}-\frac{3}{2}\right],
\label{21} \end{eqnarray}
where the sum is taken over all possible particles.
The mass parameters in the potential fulfill the following boundary
conditions at the GUT scale:
\begin{eqnarray}
  m_1^2=m^2_2 & =& \mu^2_0+m_0^2 \mbox{\hspace{3mm} and}  \nn
        m^2_3 & = & B\mu_0m_0,    \label{higgs_bound}
\end{eqnarray}
  where $\mu_0$ is the value of $\mu$ at the GUT scale.
The minimization conditions
$$\frac{\partial V}{\partial\psi_1}=0, \quad
\frac{\partial V}{\partial\psi_2}=0$$
with $\psi_{1,2}={\bf Re}H^0_{1,2}$ yield:
\beq
2m_1^2&=&2m_3^2\tan \beta - M_Z^2\cos 2\beta - 2\Sigma_1 \label{2m1} \\
2m_2^2&=&2m_3^2\cot \beta + M_Z^2\cos 2\beta - 2\Sigma_2, \label{2m2}
\eeq
where $\Sigma_1\equiv \frac{\partial \Delta V}{\partial\psi_1} $ and
$\Sigma_2\equiv \frac{\partial \Delta V}{\partial\psi_2} $
are the one-loop corrections\cite{loopewbr}:
\begin{eqnarray}
\Sigma_1   & = & -\frac{1}{32\pi^2}
\sum_i (-1)^{2J_i}(2J_i+1)\frac{1}{\psi_1}\frac{\partial m_i^2}
{\partial \psi_1}f(m_i^2) \label{sigma1} \\
\Sigma_2 & = & -\frac{1}{32\pi^2}
\sum_i (-1)^{2J_i}(2J_i+1)\frac{1}{\psi_2}\frac{\partial m_i^2}
{\partial \psi_2}f(m_i^2) \label{sigma2}
\end{eqnarray}
and  the function $f$\footnote{This definition differs by a factor 2 from
the one of Ellis et al.~\cite{erz}} is defined as:
\begin{eqnarray}
f(m^2) & = & m^2\left(\log\frac{m^2}{Q^2}-1\right)
\end{eqnarray}
The Higgs masses can now be calculated including all 1-loop corrections
\cite{erz,berz,drno,kz92,eqz,cpr}.
%%The matrix of second derivatives with respect to $\phi_i ={\bf Im}
%%H^0_i$ has the form:
%
%%The matrix of second derivatives with respect to $\psi_i ={\bf Re}
%%H^0_i$ has the form:
%
%%More general formulae are given in Ref.~\cite{kaza1}.
%%The final values for $m_h$ and $m_A$ are shown
%%in fig.\ref{mh} and \ref{ma}
%The Higgs masses corresponding to this one loop potential
%are~\cite{erz,berz,drno,kz92,eqz,cpr}:
%
%
\section{Comparision of the MSSM with experimental Data}
 In this section the various low energy GUT predictions are
compared with data. The most  restrictive constraints are
the coupling constant unification and the requirement that the
unification scale has to be above $10^{15}$ GeV from the proton
lifetime limits, assuming decay via s-channel exchange of heavy
gauge bosons. They exclude the SM~\cite{ekn2,abf,lalu} as well
as many other models~\cite{abf,abfI,yana}. The only model known to
be able to fulfill all constraints simultaneously is the MSSM.
In the following we shortly summarize the experimental inputs
and then discuss the fit results.

\subsection{Coupling Constant Unification}
The three coupling constants of the known symmetry groups are:
\bq\label{SMcoup}{\matrix{
\alpha_1&=&(5/3)g^{\prime2}/(4\pi)&=&5\alpha/(3\cos^2\theta_W)\cr
\alpha_2&=&\hfill g^2/(4\pi)&=&\alpha/\sin^2\theta_W\hfill\cr
\alpha_3&=&\hfill g_s^2/(4\pi)\cr}}
\eq%
where $g'~,g$ and $g_s$ are the $U(1)$, $SU(2)$ and $SU(3)$ coupling
constants.

 The couplings, when defined as
effective values including loop corrections in
the gauge boson propagators,
become energy dependent (``running'').  A running coupling requires
the specification of a renormalization prescription, for which
  the modified minimal subtraction ($\overline{MS}$)
scheme~\cite{msbar} is used.

In this scheme the world averaged values of the couplings at the
Z$^0$ energy are obtained from a fit to the LEP data~\cite{LEP}, $M_W$
\cite{PDB} and \mt\cite{CDF,D0}:
\begin{eqnarray}
  \label{worave}
  \alpha^{-1}(M_Z)             & = & 128.0\pm0.1\\
  \sin^2\theta_{\overline{MS}} & = & 0.2319\pm0.0004\\
  \alpha_3                     & = & 0.125\pm0.005.
\end{eqnarray}
The value of $\alpha^{-1}(M_Z)$ was updated from Ref. \cite{dfs} by using
  new data on the hadronic vacuum polarization\cite{EJ95}. %\cite{Schw}.
 For SUSY models, the dimensional reduction $\overline{DR}$
scheme is a more appropriate renormalization scheme~\cite{akt}.
In this scheme  all thresholds are  treated by simple step
approximations and unification occurs  if all three
$\aii(\mu)$ meet exactly at one point. This crossing point
corresponds to the mass of the heavy gauge bosons.
The $\overline{MS}$ and $\overline{DR}$ couplings
differ by a small offset
\bq\label{MSDR}{{1\over\alpha_i^{\overline{DR}}}=
{1\over\alpha_i^{\overline{MS}}}-{C_i\over\strut12\pi}
}\eq
where the $C_i$ are the quadratic Casimir coefficients of the
group ($C_i=N$ for SU($N$) and 0 for U(1)
so $\alpha_1$ stays the same).
 Throughout the following, we use the
$\overline{DR}$ scheme for the MSSM.
\subsection{\mz from Electroweak Symmetry Breaking}
Radiative corrections can trigger
spontanous symmetry breaking in the electroweak sector, if
the minimum is obtained for non-zero vacuum expectation
values of the fields.
Solving $\mz$ from the minimization conditions
(eqns. \ref{2m1} and \ref{2m2}) yields:
\bq
\label{defmz}
\frac{\mz^2}{2}=\frac{m_1^2+\Sigma _1-(m_2^2+\Sigma _2) \tan^2\beta}
{\tan^2\beta-1}
\eq
where the $\Sigma_1$ and $\Sigma_2$ are defined in
eqns.~(\ref{sigma1},\ref{sigma2}).
This condition  determines the value  of $\mu_0$ for a given value
of $\mze$ and $\mha$, as follows from the boundary values
 of $m_1$ and $m_2$ (\ref{higgs_bound}).
Furthermore one can express $m_3^2=B\mu\mze$  as function of $\tb$,
so one can exchange the parameter $B$ with $\tb$, as will be done in the
following.
\subsection{Yukawa Coupling~Constant Unification}
\label{masses}
 The masses of top, bottom and $\tau$ can be obtained from the
 low energy values of the running yukawa couplings as shown in
 eq. (\ref{yukawastbtau}). The requirement of bottom-tau Yukawa
 coupling unification strongly restricts the  possible solutions in
 the $\mt$ versus $\tb$ plane, as discussed by many
 groups     %~\cite{rrb92,ir92},~\cite{Ross1} --
 \cite{Ross1,bbog,lanpol,bmaskln,cpr,copw1,ara91}.
 %        \cite{ara91}.
The values of the running masses can be translated to
pole masses following the formulae from~\cite{runmas}.
In the MSSM the bottom mass has additional corrections from loops
involving gluinos, charginos and charged Higgs bosons~\cite{hemp,hall}.
These corrections are small
for low \tb solutions, but
become large for the high \tb values.
For the pole masses of the third generation the following values
are taken:
$M_t=179\pm 12~GeV/c^2$~\cite{CDF,D0},
$M_b=4.94\pm0.15~GeV/c^2$~\cite{PDB} and
$M_\tau=1.7771\pm0.0005~GeV/c^2$
{}~\cite{PDB}.
Since the gauge couplings are measured most precisely at $\mz$, the
Yukawa couplings were fitted  at $\mz$ too.
The pole mass  of the b-quark at $\mz$ was calculated by using the third
order QCD formula\cite{ckw93},
which leads to $M_b(\mz)=2.84\pm0.15~GeV/c^2$ for $\as(\mz)=0.125\pm0.005$;
  the error on $M_b$ includes the uncertainty from $\as$.
The running of   $M_\tau$
is much less between $M_\tau$ and $\mz$; one finds
$M_\tau(\mz) = 1.7462\pm0.0005$. The Yukawa coupling of the top quark
is always evaluated at $M_t$, since its running depends on the SUSY spectrum,
which may be splitted in particles below and above $M_t$.
\subsection{Experimental Lower Limits on SUSY Masses}
SUSY particles have not been found so far
and from the searches
at LEP one knows that the lower limit on the
charged leptons and charginos is
about half the $Z^0$ ~mass (45 GeV)~\cite{PDB}
and the Higgs mass has to be above
60 GeV~\cite{higgslim,sopczak}. The lower limit on
the lightest neutralino is 18.4 GeV~\cite{PDB},
while the sneutrinos have to
be above 41 GeV~\cite{PDB}.
  These limits require  minimal values for the
SUSY mass parameters.
There exist also limits on squark and gluino
masses from the hadron colliders~\cite{PDB}, but these
limits depend on the assumed decay modes.
Furthermore, if one takes the limits given above
into account, the  constraints from the limits on all other
particles are usually fulfilled, so they
do not provide additional reductions of  the
parameter space in case of the {\it minimal} SUSY model.
\subsection{Branching Ratio $BR(b\to s \gamma)$}
The branching ratio \Bbsg has been measured by the CLEO
collaboration \cite{cleo94} to be:
$BR(b\to s \gamma)=2.32\pm0.67\times10^{-4}$.

In the MSSM this flavour changing neutral current (FCNC)
receives, in addition to the SM $W-t$ loop, contributions from
$H^\pm-t$ and $\tilde{\chi}^\pm -t$ loops.
The $\tilde{g}-\tilde{q}$ and $\tilde{\chi}^0 -t$ loops, which are
expected to be much smaller, have been neglected\cite{borz,bsgamm3}.
The chargino contribution, which becomes large
for large $\tb$ and small chargino masses,  depends
sensitively on the splitting
of the two stop masses; therefore it is important to
diagonalize the matrix without approximations.

The theoretical prediction depends on the renormalization scale~\cite{burasb}.
Varying the scale between $m_b/2$ and $2m_b$ leads to a theoretical
uncertainity $\sigma_{th.}=0.6\times10^{-4}$, which is added in quadrature
to the experimental error. The fit prefers scales close to the upper limit,
so the analysis was done with $2m_b$ as renormalization scale.

Within  the MSSM   the following ratio has been
calculated \cite{bsgamma,bsgamm3}:
\begin{equation}
\frac{BR(b\to s\gamma)}{BR(b\to c e
\bar{\nu})}=\frac{|V_{ts}^*V_{tb}|^2}{|V_{cb}|^2}
K_{NLO}^{QCD}\frac{6\alpha}{\pi}
\frac{\left[\eta^{16/23}A_\gamma+\frac{8}{3}(\eta^{14/23}
-\eta^{16/23})A_g+C\right]^2}{I(m_c/m_b)
[1-(2/3\pi)\alpha_s(m_b)f(m_c/m_b)]},
\end{equation} where
\begin{eqnarray}
 \eta&=&\alpha_s(M_W)/\alpha_s(m_b)\\
 f(m_c/m_b)&=&2.41.
\end{eqnarray}
Here $ f(m_c/m_b)$ represents corrections
from {\em leading order} QCD to the known semileptonic
$b\to c e \bar{\nu}$  decay rate, while   the ratio of
masses of c- and b-quarks is taken to be
  $m_c/m_b=0.316$. The ratio of CKM matrix elements
  $\frac{|V_{ts}^*V_{tb}|^2}{|V_{cb}|^2}=0.95$ was
taken from Buras et al.
{}~\cite{burasb} and the factor $K_{NLO}^{QCD}=0.83$ for
the {\em next leading order } QCD-Corrections
from Ali et al.~\cite{alibsg}.

Comparing these formulae with the experimental
results leads to significant constraints on the
parameter space, especially at large values
of $\tb$, as discussed by many groups
\cite{bop,roskane,bsgamm1,bsgamm2}.

\subsection{Dark Matter Constraint }\label{s67}
Abundant evidence for the existence
of non-relativistic, neutral, non-baryonic
dark matter exists in our universe\cite{borner,kolb}.
The lightest supersymmetric particle (LSP) is supposedly stable
and would be  an ideal candidate for dark  matter.

The present lifetime of the universe is at least $10^{10}$ years,
which implies an upper limit on the expansion rate
 and correspondingly on the total relic abundance.
Assuming $h_0>0.4$ one finds that  the contribution of
 each relic particle species $\chi$  has to obey~\cite{kolb}:
  \beq\Omega_\chi h^2_0<1,\eeq
where $\Omega_\chi h^2$ is the ratio of the relic particle density
of particle $\chi$ and the critical density, which
overcloses the universe.
This bound can only be met, if
most of the LSP's   annihilated into
fermion-antifermion pairs, which in turn would
annihilation into  photons again.

Since the neutralinos are mixtures of gauginos and
higgsinos, the annihilation can occur both, via
s-channel exchange of the $Z^0$ and Higgs bosons and
 t-channel exchange of a scalar particle, like a selectron \cite{relic}.
This constrains the parameter space, as discussed by
many groups\cite{relictst,roskane,rosdm,bop}.
The size of the Higgsino component
depends on   the relative sizes of the elements
in the mixing matrix (eq. \ref{neutmix}), especially
on the mixing angle $\tb$ and the size of the
parameter $\mu$ in comparison to $M_1\approx
0.4m_{1/2}$ and $M_2\approx 0.8 m_{1/2}$.
This mixing becomes large for the SO(10) type solutions,
in which case the parameters can alway be tuned such,
that the relic density is low enough.

However, for low $\tb$ values
the mixing is very small due to the large value
of $\mu$ required from electroweak symmetry breaking
and one finds that the
lightest scalars have to be below a few 100 GeV in that case,
as will be discussed below.
The relic density was computed from the formulae
by Drees and Nojiri~\cite{drno} and from the more approximate
formulae by Ellis et al.~\cite{lspdark}.
They typically agree within a factor two, which is
satisfactory and good enough, since  the relic density
is such a steep function of the parameters for low $\tb$, that the
excluded regions are hardly changed by a factor two
uncertainty.

\subsection{Fit Method}
The fit method has been described in detail before~\cite{bek} for the low
$\tb$ region. In that case the analytical solutions for the SUSY masses
could be found and one had to integrate only four RGE ($Y_t$ and $3\alpha_i$)
numerically. For large $\tb$ values all  25 RGE's of section
 \ref{rge} have to be integrated
simultaneously. As a check, this integration was performed for low $\tb$ values
too and found to be in good agreement with the results using the analytical
solutions for the masses. In the present analysis the following $\chi^2$
definition is used:
\begin{eqnarray}
 \chi^2 & = &
{\sum_{i=1}^3\frac{(\aii(\mz)-\alpha^{-1}_{MSSM_i}(\mz))^2}
{\sigma_i^2}}
          \nn
 & &+\frac{(\mz-91.18)^2}{\sigma_Z^2}     \nn
 & &+{\frac{(\mt - 174)^2}{\sigma_t^2}}          \nn
 & &+\frac{(\mb-4.98)^2}{\sigma_b^2}       \nn
 & &+\frac{(m_\tau-1.7771)^2}{\sigma_\tau^2}       \nn
 & &+{\frac{(Br(b\to s\gamma)-2.32 \times 10^{-4})^2}
     {\sigma(b\to s\gamma)^2}} \nn
 & &+{\frac{(\Omega h^2-1)^2}{\sigma^2_\Omega}} ~ (for ~\Omega h^2 > 1) \nn
 & &+{\frac{(\tilde{M}-\tilde{M}_{exp})^2}{\sigma_{\tilde{M}}^2}}
 ~ {(for~\tilde{M} > \tilde{M}_{exp})}  \nn
 & &+{\frac{(\tilde{m}_{LSP}-\tilde{m}_{\chi})^2}{\sigma_{LSP}^2}}
 ~ {(for~ \tilde{m}_{LSP}~charged )}\nn.                  \label{chi2}
\end{eqnarray}
The first six terms are used to enforce gauge coupling unification, electroweak
symmetry breaking and $b-\tau$ Yukawa coupling unification, respectively.
The following two terms impose the constraints from \bsg~ and
the relic density, while the
 last terms require the SUSY masses to be above the
experimental lower limits and  the lightest sypersymmetric
particle (LSP) to be  a neutralino, since a charged stable LSP would have
been observed.
The input and fitted output variables have been summarized
in table \ref{t1}.
\begin{table}[htb]
\begin{center}
\begin{tabular}{|c||c||c|c|}
\hline
                     &&
                       \multicolumn{2}{|c|}{Fit parameters}      \\
\cline{3-4}
\rb{exp.~input data} & \rb{$\Rightarrow$} & low $\tb$
                                               & high $\tb$     \\
\hline
$\alpha_1,\alpha_2,\alpha_3$ &   & \mgut,~\agut  & \mgut,~\agut\\
%\cline{1-1}\cline{3-4}
\mt                          &   & $Y_t^0,~Y_b^0=Y_\tau^0$ &
                                   $Y_t^0=Y_b^0=Y_\tau^0$ \\
%\cline{1-1}\cline{3-4}
\mb              &\rb{ minimize}& \mze,\mha & \mze,\mha  \\
%\cline{1-1}\cline{3-4}
 $m_\tau$        &\rb{ $\chi^2$} & \tb &    \tb      \\
%\cline{1-1}\cline{3-4}
\mz                          &   & $\mu$ &  $\mu$      \\
%\cline{1-1}\cline{3-4}
\bsg                         &   & $(A_0)$ & $A_0$ \\
 $\tau_{universe}$           &   &             &               \\
\hline
\end{tabular} \end{center}
\caption[]{\label{t1}Summary of fit input and output variables. For
the low $\tb$ scenario the parameter $A_0$ is not very relevant as
indicated by the brackets. For large $\tb$ $\tau_{universe}$ does not
yield any constraints (see text).}
\end{table}

%\begin{table}[bth]
%\renewcommand{\rb}[1]{\raisebox{1.75ex}[-1.75ex]{#1}}
%\begin{center}
%\begin{tabular}{|c||c||c|}
%\hline
%Input  & $\Rightarrow$& Output\\
%\hline
%$\alpha_1,\alpha_2,\alpha_3$& & \mgut,~\agut\\
%
%\mt        & &\mze,\mha\\
%
%\mb   &\rb{ Minimize} & \tb \\
%
%$m_\tau$ &\rb{ $\chi^2$} & $A_0 $\\
%
%\mz & &  $\mu$  \\
%
%\bsg & &  \\
%$\tau_{universe}$ & &  \\
%\hline
%\end{tabular} \end{center}
%\caption[]{\label{t1}Summary of fit input and output variables.}
%\end{table}
%
%
\begin{table}[h]
\vspace*{-1.7cm}
\renewcommand{\arraystretch}{1.30}
\renewcommand{\rb}[1]{\raisebox{1.75ex}[-1.75ex]{#1}}
\begin{center}
\begin{tabular}{|c|r|r|}
\hline
 \multicolumn{3}{|c|}{ Fitted SUSY parameters }                       \\
\hline
\hline
Symbol & \makebox[3.0cm]{\bf{low $\tb$}} & \makebox[3.0cm]{\bf{high $\tb$}}\\
\hline
 $m_0$     	&  200 		&  600                 	\\
\hline
 $m_{1/2}$ 	&  270 		&  70			\\
\hline
 $\mu(0)$     	& -1084 	& -196			\\
\hline
 $\mu(\mz)$  	&  -546         & -140	\\
\hline
 $\tan\beta$	& 1.71 		& 45.5 			\\
\hline
 $Y_t(\mt)$  	&  0.0080	& 0.0057 		\\
\hline
 $Y_t(0)$  	& 0.0416	& 0.0020		\\
\hline
 $Y_b(0)$  	& 0.1188E-05	& 0.0015		\\
\hline
$M_t^{pole}$ 	& 177 		& 174			\\
\hline
 $m_t^{running}$   	& 168		& 165   		\\
\hline
 $1/\alpha_\rG$	& 24.8		& 24.2			\\
\hline
 $\MG$  	& $1.6\;10^{16}$& $2.4\;10^{16}$ 	\\
\hline
\hline
 $A(0)\mze$  	& 0 & 536 	\\
\hline
 $A_t(\mz)\mze$  	&    -446             & -41	\\
\hline
 $A_b(\mz)\mze$  	&    -886            & -33	\\
\hline
 $A_\tau(\mz)\mze$  &    -546            & 231	\\
\hline
 $m_1(\mz)$  &    612           & 150	\\
\hline
 $m_2(\mz)$  &    262           & -131	\\
\hline
\end{tabular} \end{center}
\caption[]{\label{t2a}Values of the fitted SUSY parameters
             for low and high $\tb$.
             The scale is  either $\mz$, $\mt$,  or $\rG$, as indicated
             in the first column by ($\mz$), ($\mt$) or (0), respectively.
             The SUSY mass spectrum corresponding to these parameters
             is given in table \ref{t2}.}
\end{table}
\begin{table}[h]
\vspace*{-1.7cm}
\renewcommand{\arraystretch}{1.30}
\renewcommand{\rb}[1]{\raisebox{1.75ex}[-1.75ex]{#1}}
\begin{center}
\begin{tabular}{|c|r|r|}
\hline
 \multicolumn{3}{|c|}{SUSY masses in [GeV]}  	        \\
\hline
\hline
Symbol & \makebox[3.0cm]{\bf{low $\tb$}} & \makebox[3.0cm]{\bf{high $\tb$}}\\
\hline
\hline
  $\tilde{\chi}^0_1(\tilde{B})$   	&  116 	& 25		\\
\hline
  $\tilde{\chi}^0_2(\tilde{W}^3)$        	&  231	& 46		\\
\hline
  $\tilde{\chi}^{\pm}_1(\tilde{W}^\pm)$    	&  231	& 46		\\
\hline
  $\tilde{g}$                  	&  658  & 191		\\
\hline  \hline
  $\tilde{e}_L$      		&  278 	& 604		\\
\hline
  $\tilde{e}_R$    		&  228  & 602		\\
\hline
  $\tilde{\nu}_L$     		&  273  & 599		\\
\hline  \hline
  $\tilde{q}_L$   		&  628	& 622		\\
\hline
  $\tilde{q}_R$   		&  605  & 620		\\
\hline
  $\tilde{\tau}_1$    		&  227	& 423		\\
\hline
  $\tilde{\tau}_2$   		&  228	& 525		\\
\hline
  $\tilde{b}_1$       		&  560	& 352		\\
\hline
  $\tilde{b}_2$   		&  604	& 426		\\
\hline
  $\tilde{t}_1$    		&  477	& 394		\\
\hline
  $\tilde{t}_2$  		&  582	& 413		\\
\hline        \hline
  $ \tilde{\chi}^0_3(\tilde{H}_1)$  	&  562  & (-)163		\\
\hline
  $ \tilde{\chi}^0_4(\tilde{H}_2)$   	& (-) 571	& 178		\\
\hline
  $\tilde{\chi}^{\pm}_2(\tilde{H}^{\pm})$& 569	& 185		\\
\hline   \hline
  $       h $    		&  81  	& 105		\\
\hline
  $       H $   		&  739	& 177		\\
\hline
  $       A $   		&  734 	& 182		\\
\hline
  $       H ^{\pm}$    		&  738	& 200		\\
\hline  \hline
  $\Omega h^2$                  &  0.42     & 0.025         \\
\hline
  Br(\bsg)			& $2.87\;10^{-4}$ & $2.36\;10^{-4}$ \\
\hline
  LSP$\rightarrow |\tilde{B}>$	& 0.9973& 0.9141\\
\hline
  LSP$\rightarrow |\tilde{W}^3>$	& 0.0360& -0.1354\\
\hline
  LSP$\rightarrow |\tilde{H}_1^0>$	& -0.0593& -0.3770\\
\hline
  LSP$\rightarrow |\tilde{H}_2^0>$	& 0.0252&-0.0635\\
\hline
\end{tabular} \end{center}
\caption[]{\label{t2}Values of the  SUSY
            mass spectra for the low and high $\tb$ solutions, given in
            table \ref{t2a}. The (-) in front of the neutralinos indicates
            that it is a CP-odd state. The LSP is a linear combination of the
            gaugino and Higgsino components, as indicated by the last four
rows.
            Note the much larger Higgsino component of the LSP for large
            $\tb$, which leads to a small relic density. }
\end{table}
\section{Results}
\begin{figure}
 \begin{center}
  \leavevmode
  \epsfxsize=15cm
  \epsffile{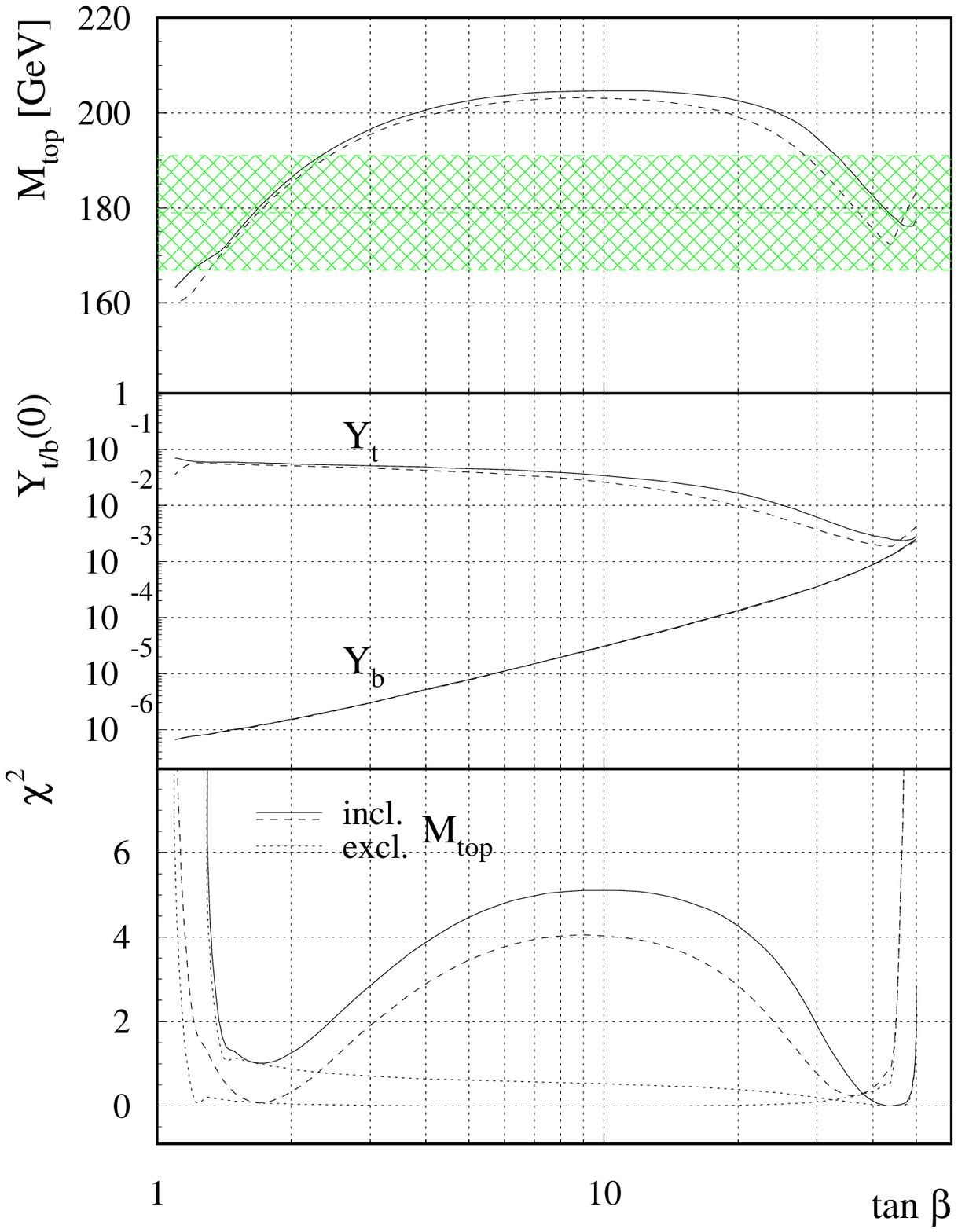}
\end{center}
\caption[]{\label{\figI}The top quark mass as function of $\tb$ (top)
for values of $\mze,\mha$ optimized for low and high $\tb$, as indicated
by the dashed and solid lines, respectively.
The middle part shows the corresponding values of the Yukawa
coupling at the GUT scale and the lower part the obtained
$\chi^2$ values. If the top constraint ($\mt=179\pm12$, horizontal band)
is not applied, all values of $\tb$ between 1.2 and 50 are allowed
(thin dotted lines at the bottom), but if the top mass
is constrained to the experimental value, only the regions
$1<\tb<3$ and $15<\tb<50$ are allowed.}
%\label{chi2plot}
\end{figure}

The requirement of bottom-tau Yukawa coupling unification strongly
restricts the  possible solutions in the $\mt$ versus $\tb$ plane,
as discussed before.
With the top mass measured by
the CDF and D0-Collaborations~\cite{CDF,D0}
 only two regions of $\tb$ give an acceptable
$\chi^2$  fit, as shown in the  bottom part of fig. \ref{\figI}
for two values of the SUSY scales $\mze,\mha$, which are optimized for
the low and high $\tb$ range, respectively,
as will be discussed below.
 The curves at the top show the solution for $\mt$
as function of $\tb$ in comparison with the
experimental value of $\mt=179\pm 12$ GeV. The  $\mt$ predictions
 were obtained by imposing gauge coupling unification
and electroweak symmetry breaking for each value of $\tb$,
which allows a determination of $\mu, ~\agut$, and $\mgut$
from the fit for the given choice
of $\mze,\mha$. Note that the results do not depend very much on this
choice.
The influence of the large corrections to $\mb$ at large values
of $\tb$ and the constraints from Br(\bsg) will be discussed below.

\begin{figure}
 \begin{center}
  \leavevmode
  \epsfxsize=15cm
  \epsffile{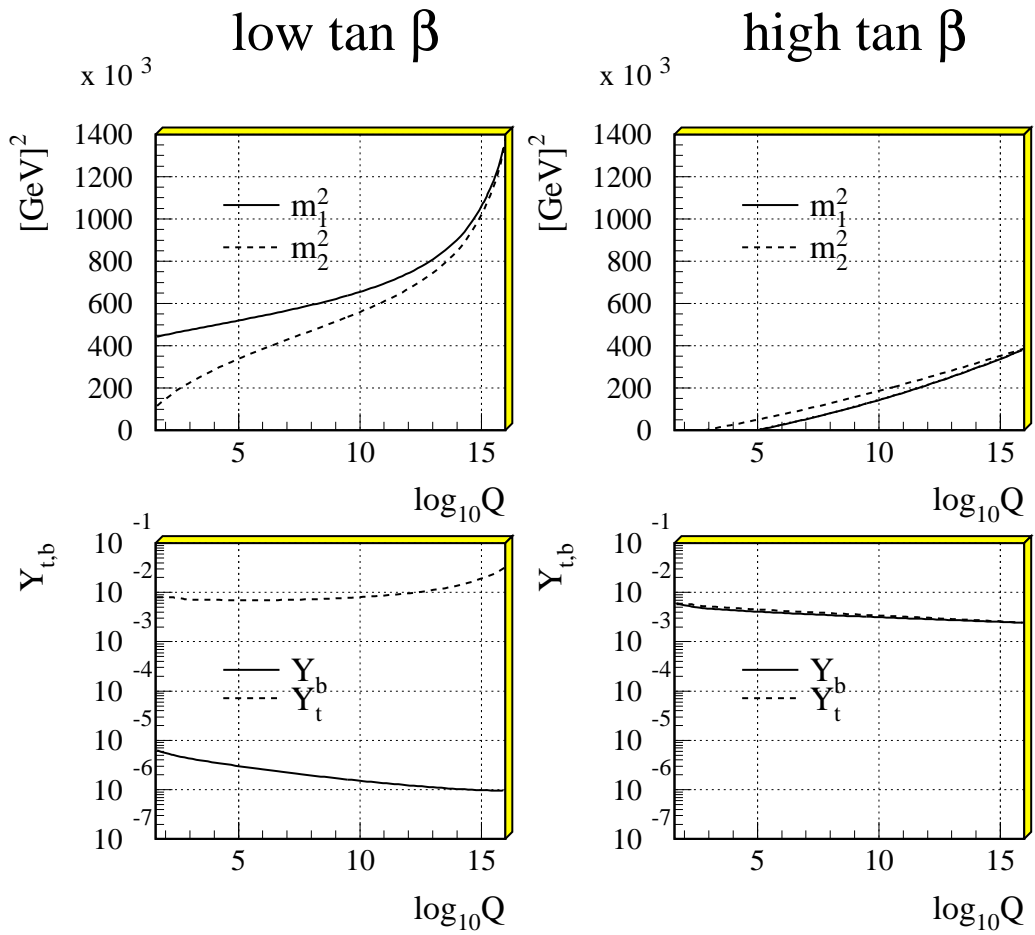}
\end{center}
\caption[]{\label{\figII}The running of the parameters $m_1$ and $m_2$ in
         the Higgs potential (top) and Yukawacouplings of top and bottom
         quarks (bottom).}
\end{figure}

The best $\chi^2$ is obtained
for $\tb=1.7$ and $\tb=46$, respectively. They correspond  to
 solutions where $Y_t>>Y_b$ and $Y_t\approx Y_b$,
as shown in the middle part of fig. \ref{\figI}.	The latter solution
is the one typically expected for the SO(10) symmetry, in which
the up and down type quarks as well as leptons
belong to the same multiplet, while
the first solution corresponds to b-tau unification only,
as expected for the minimal SU(5) symmetry.
In SO(10) exact top-bottom Yukawa unification is difficult,
mainly because of the requirement of radiative electroweak symmetry
breaking, since in that case  both, the mass parameters
in the Higgs potential ($m_1$ and $m_2$) as well as the Yukawa couplings,
stay similar at all energies, as shown in fig. \ref{\figII}.
Since eq. (\ref{defmz}) for $\mz$ can be written as
\beq
    \tan^2\beta & = & \frac{m_1^2 + \frac{1}{2}M_Z^2}
                           {m_2^2 + \frac{1}{2}M_Z^2}, \label{tanbet}
\eeq
one observes immediately that large values of $\tb$ cannot be obtained
if $m_1^2 \approx m_2^2$. For small $\tb$
$m_1^2$ and $m_2^2$ are sufficiently different due to the large
difference between the top and bottom Yukawa couplings
(see fig. \ref{\figII}). Since the  large $\tb$ solutions
require a judicious finetuning in case of exact unification
at the GUT scale, a small non-unification is assumed, which
could   result from threshold effects or running of the
parameters between the Planck scale and  the GUT scale.
E.g. if the SO(10) symmetry would be broken into SU(5) below the Planck
scale, but well above the GUT scale, the top Yukawa coupling
could be  easily 20-30\% larger than the bottom Yukawa coupling,
as estimated from the SU(5) RGE. Therefore, in the following
analysis $Y_t$ is taken to be 25\% larger than $Y_b$ at the GUT scale
and a similar splitting was introduced between $m_1^2$ and $m_2^2$,
i.e. $m_1^2=1.25~m_0^2 + \mu^2$ and $m_2^2=1.0~m_0^2 + \mu^2$ at the GUT scale.
It is interesting to note that
the SU(5) RGE predicts $Y_t>Y_b$ and that indeed fits with
$Y_t<Y_b$ at the GUT scale did not converge, but with the
mentioned deviations from exact unification the fits easily converged.

\begin{figure}
 \begin{center}
  \leavevmode
  \epsfxsize=15cm
  \epsffile{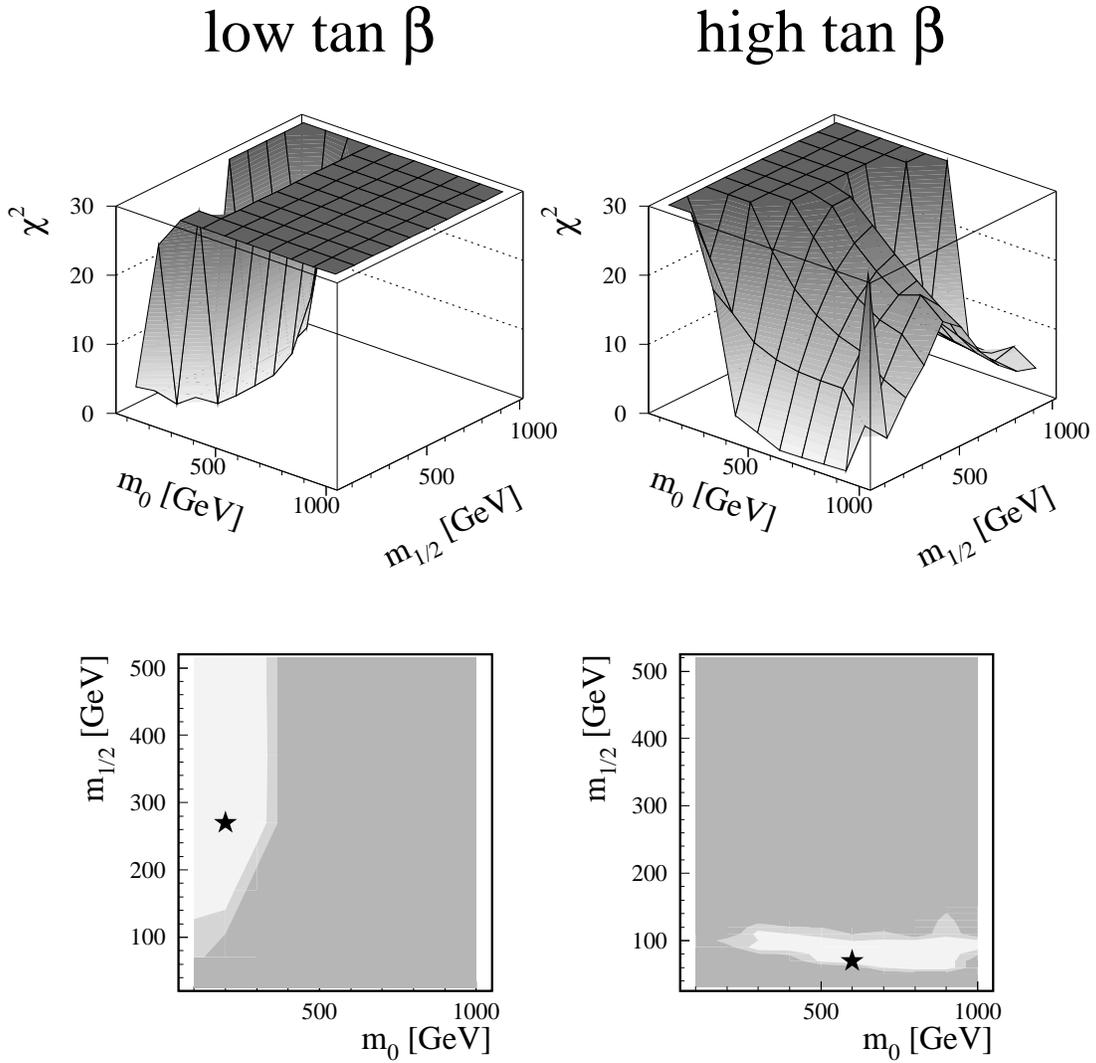}
\end{center}
\caption[]{\label{\figIII}The total $\chi^2$-distribution for
         low and high $\tb$  solutions (top) as well as the projections
         (bottom). The different shades indicate steps of $\Delta\chi^2 = 2$,
         so basically only the light shaded region is allowed.
         The stars indicate
         the optimum solution.}
\end{figure}

In fig. \ref{\figIII} the total $\chi^2$ distribution is shown as
funtion of $\mze$ and $\mha$ for the two values of $\tb$ determined
above. One observes clear minima at $\mze,\mha$ around (200,270)
and (600,70), as indicated by the stars in the projections.
The different shades correspond to   $\Delta \chi^2$ steps of 2.
Note the sharp increase in $\chi^2$, so basically only the
light shaded regions are allowed independent of the
exact $\chi^2$ cut.

\begin{figure}
 \begin{center}
  \leavevmode
  \epsfxsize=13cm
  \epsffile{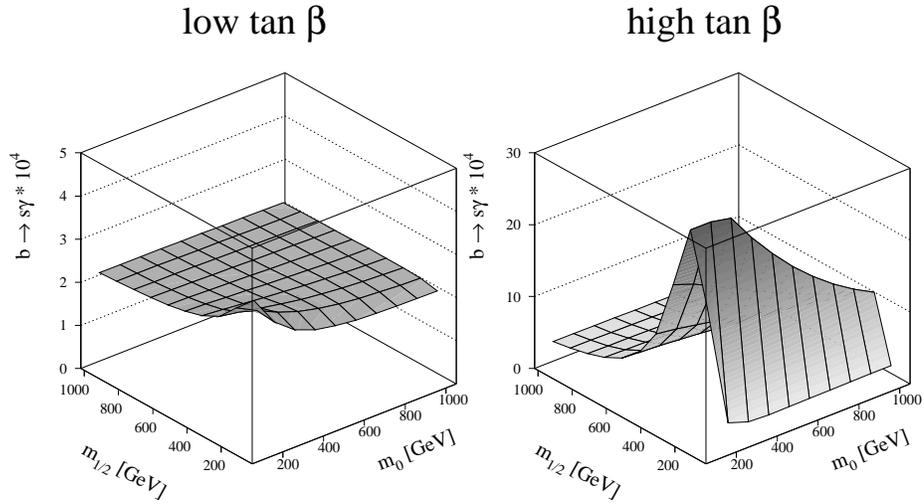}
\end{center}
\caption[]{\label{\figV}The branching ratio \bsg as function of $\mze$ and
         $\mha$. Note that for large \tb~only the region for
  $\mha < 120$ GeV yields values compatible with  experimental results.}
\end{figure}

\begin{figure}
 \begin{center}
  \leavevmode
  \epsfxsize=15cm
  \epsffile{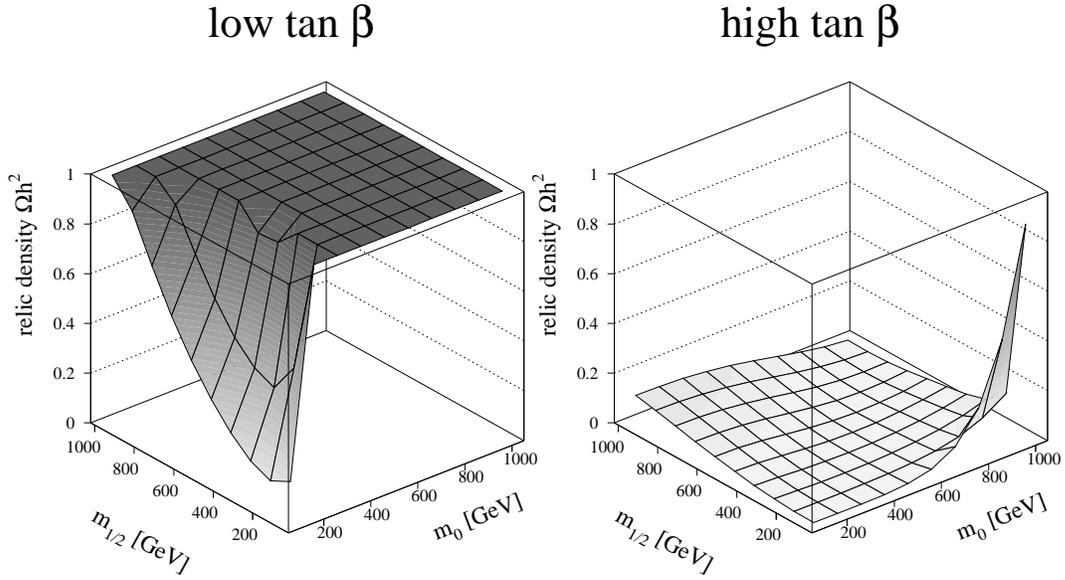}
\end{center}
\caption[]{\label{\figVI}The relic density as function of
         $\mze$ and $\mha$   for the low and high $\tb$ scenario,
respectively.}
\end{figure}

The main contributions to $\chi^2$ are different for the different
$\tb$ regimes: for large $\tb$  only small values of $\mha$
yield good fits, because of the simultaneous constraints
of \Bbsg~  and the large corrections to the b-quark mass, while
at low $\tb$ most of the $\mze,\mha$ region is
eliminated by the requirement that the relic density parameter
$\Omega h^2$ should be below one. The calculated
value of \Bbsg ~ and the relic density are shown as
function of $\mze$ and $\mha$ in figs.  \ref{\figV} and  \ref{\figVI},
respectively.

\begin{figure}
\begin{center}
  \leavevmode
  \epsfxsize=15cm
  \epsffile{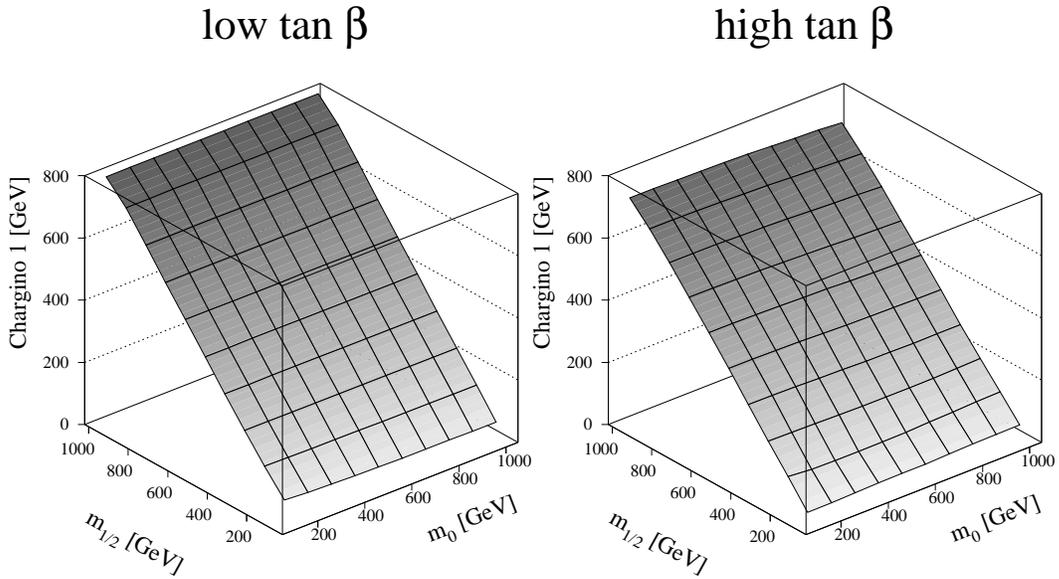}
\end{center}
\caption[]{\label{\figVII}The lightest chargino  mass as function of
         $\mze$ and $\mha$  for the low and high $\tb$ scenario, respectively.}
\end{figure}

It should be noted the mass of the lightest chargino is about
$0.7~-0.8~\mha$, as shown in fig. \ref{\figVII}.
The low value of $\mha$ for
the best fit at large $\tb$ implies a chargino mass of about
46 GeV (see table \ref{t2}), which is just above the LEP I limit and should be
detectable
at LEP II or alternatively, the large $\tb$ scenario can be excluded at
LEP II, at least the minimal version. Of course, this conclusion depends
sensitively on the \Bbsg~ value. For large $\mha$ values, the prediction
for this branching ratio is only 2 or 3 standard deviations above its
experimental value (see fig. \ref{\figV}). In non-minimal models, e.g. ones
with large splittings between $m_1$ and $m_2$ at the GUT scale, as
studied by Borzumati et al.~\cite{bop}, the prediction for this
branching ratio can be brought into agreement with experiment in the large
$\mha$ region.

\begin{figure}
 \begin{center}
  \leavevmode
  \epsfxsize=15cm
  \epsffile{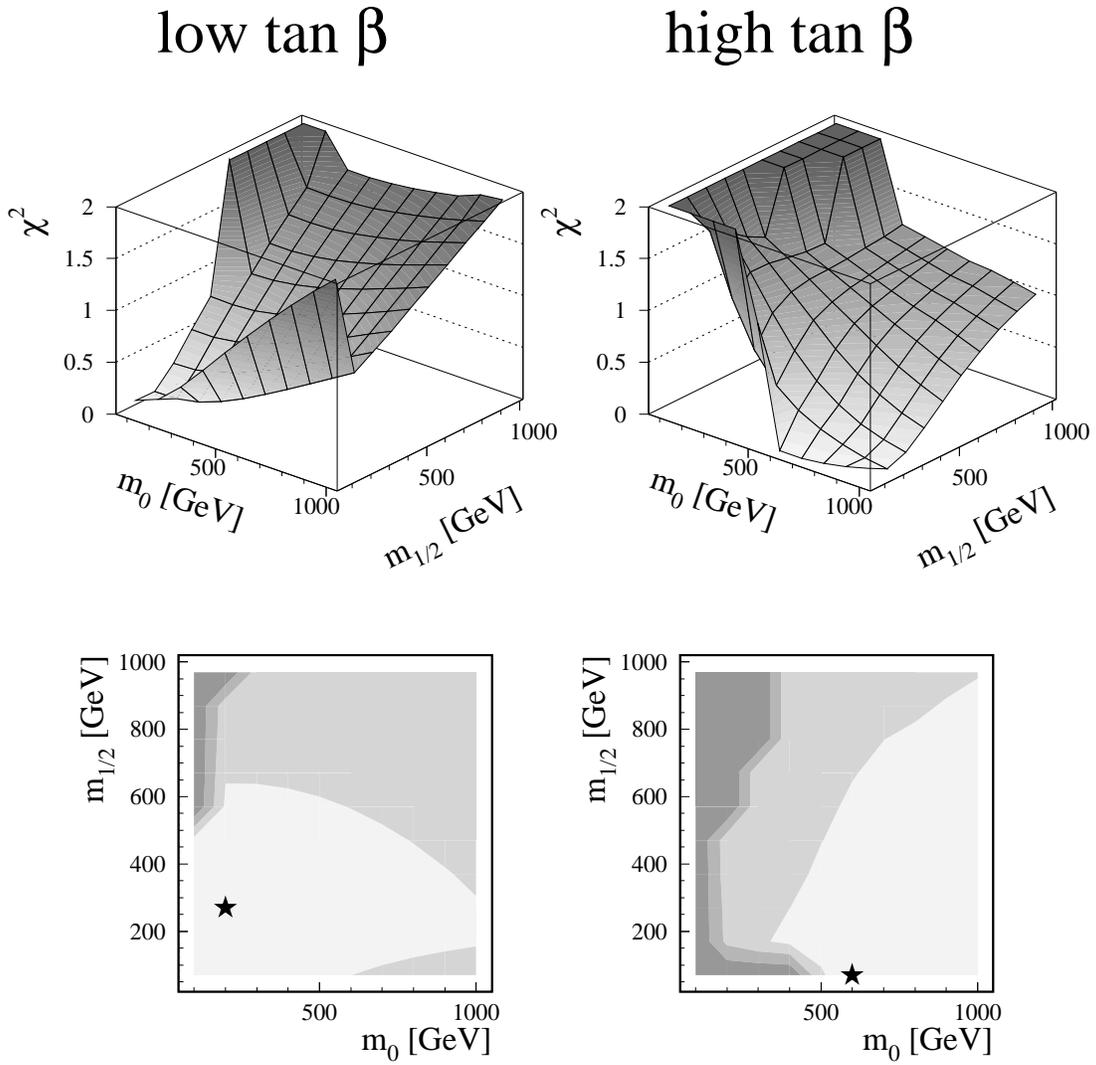}
\end{center}
\caption[]{\label{\figVIII}As  fig.  3, but only including the constraints from
         unification and electroweak symmetry breaking.}
\end{figure}

Without the contraints from \bsg and dark matter, large values
of the SUSY scale cannot be excluded, since the $\chi^2$ from gauge and
Yukawa coupling unification and electroweak symmetry breaking alone does
not exclude these regions (see fig. \ref{\figVIII}), although there is
a clear preference for the lighter SUSY scales.

\begin{figure}
 \begin{center}
  \leavevmode
  \epsfxsize=15cm
  \epsffile{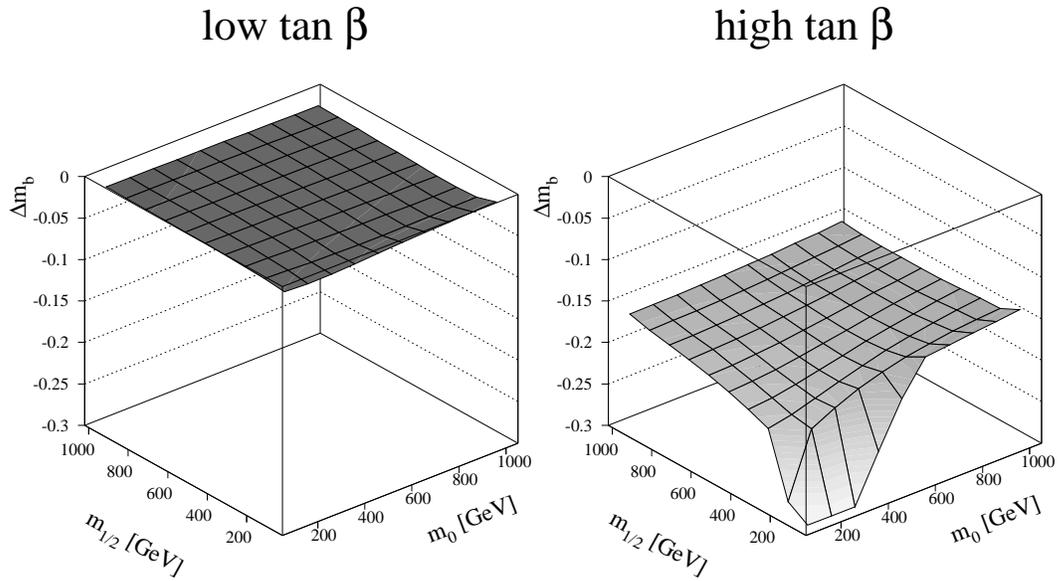}
\end{center}
\caption[]{\label{\figIX}Corrections to the bottom quark mass from gluino,
         charged Higgses and Higgsino loop contributions in the MSSM as
function
         of $\mze$ and $\mha$. Note the large negative corrections for
	 $|\mu|<0$ in this case. Positive $\mu$-values would yield a large positive
         contribution, which excludes bottom-$\tau$  unification for  most
         of the parameter space.}
\end{figure}

\begin{figure}
 \begin{center}
  \leavevmode
  \epsfxsize=15cm
  \epsffile{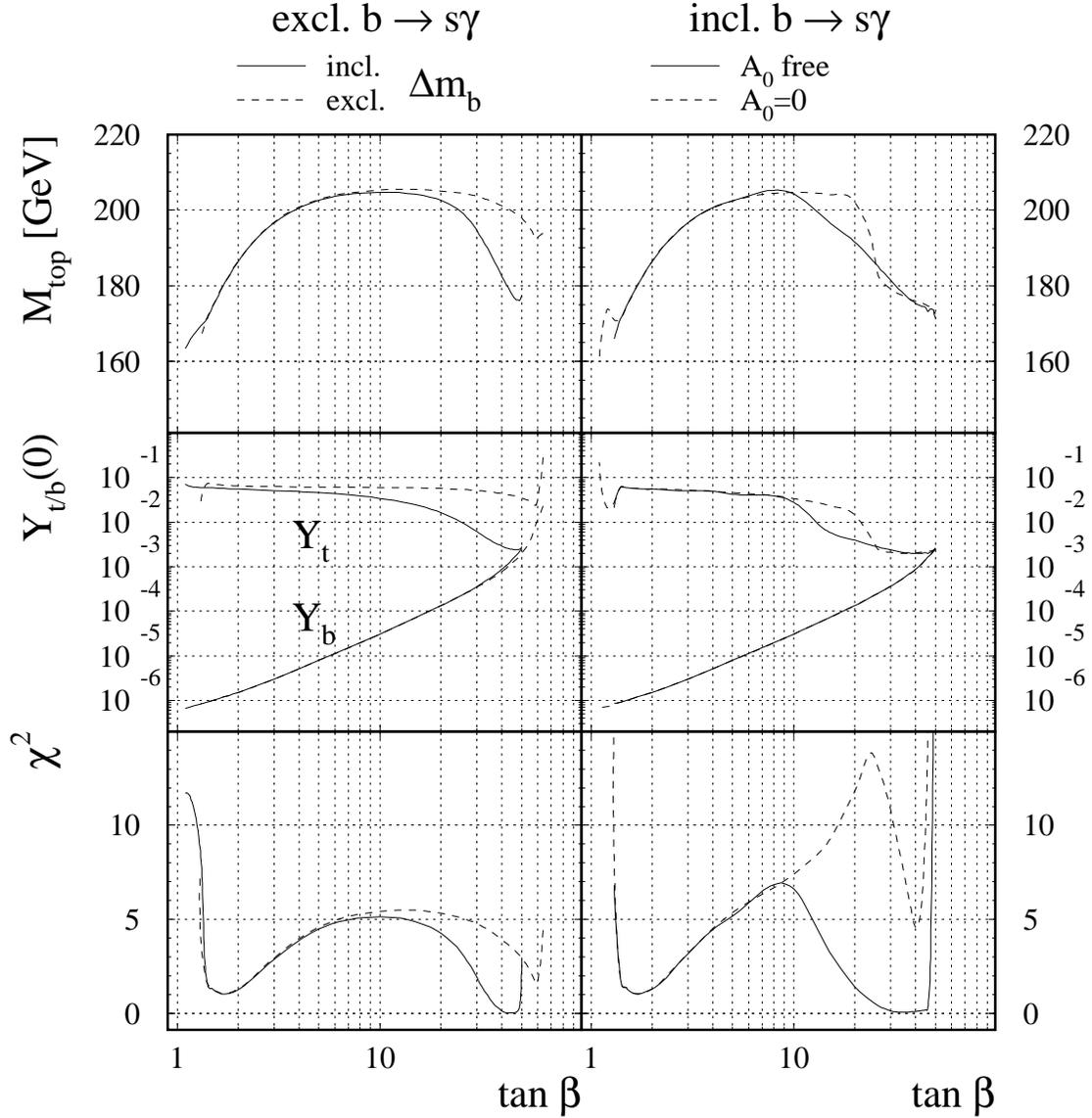}
\end{center}
\caption[]{\label{\figX}The top mass as function of $\tb$ for $\mze=600$
         and $\mha=70$ GeV. The various curves show the influence of the
         $\Delta m_b$ corrections, the \bsg branching ratio and the
         trilinear couplings of the third generation ($A_t=A_b=A_\tau=A_0$)
         at the GUT scale. These effects are only important for the
         higher values of $\tb$.}
\end{figure}

\begin{figure}
\begin{center}
  \leavevmode
  \epsfxsize=15cm
  \epsffile{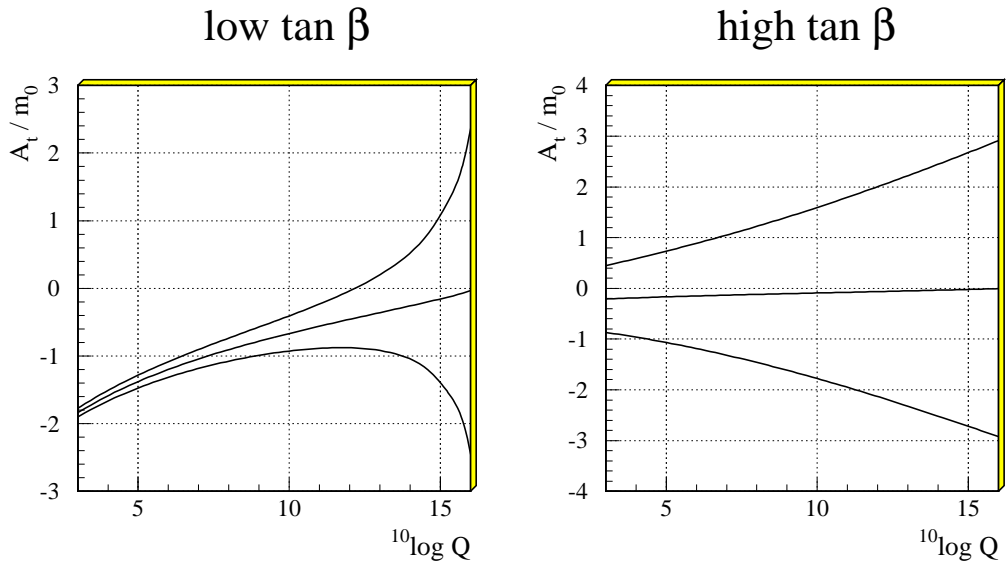}
\end{center}
\caption[]{\label{\figXa}The running of $A_t$.
 The values of ($\mze,\mha$) were choosen to be (200,270) and (600,70)
GeV  for the low and high $\tb$ scenario, respectively.
However, the {\it fixed point} behaviour is found for other values
of $\mze,\mha$ too:
at low values of $\tb$ a strong convergence to a single value is found,
while for high values this tendency is much less pronounced.}
\end{figure}

 As mentioned in the
previous section , at large $\tb$ the b-mass has large, but finite
corrections in the MSSM, as shown in fig. \ref{\figIX}. Both, \Bbsg and
$\Delta m_b$ are sensitive functions of the mixing in the quark sector,
given by the off-diagonal terms in the mass matrices
(eqns. \ref{stopmat}-\ref{staumat}). Fitting both values simultaneously
requires
the trilinear coupling $A_0$ at the GUT scale to be non-zero, as shown in
fig. \ref{\figX}: the $\chi^2$ for large $\tb$ and $A_0 = 0$ is much
worse than for  fits, in which $A_0$ is left free. The influence of
$\Delta m_b$ on the $m_t$ versus $\tb$ solution is shown too on the left side.
Note that the $\Delta m_b$ corrections improve the fit at the high $\tb$
values. For the low $\tb$ scenario the trilinear couplings were found
to play a negligible role: varying them between $\pm 3\mze$ did not change
the results significantly, since
  $A_t$ shows a {\it fixed point}
behaviour in this case: its value at $\mz$ is practically independent of the
starting value at the GUT scale, as shown in fig.  \ref{\figXa}.

\begin{figure}
\begin{center}
  \leavevmode
  \epsfxsize=15cm
  \epsffile{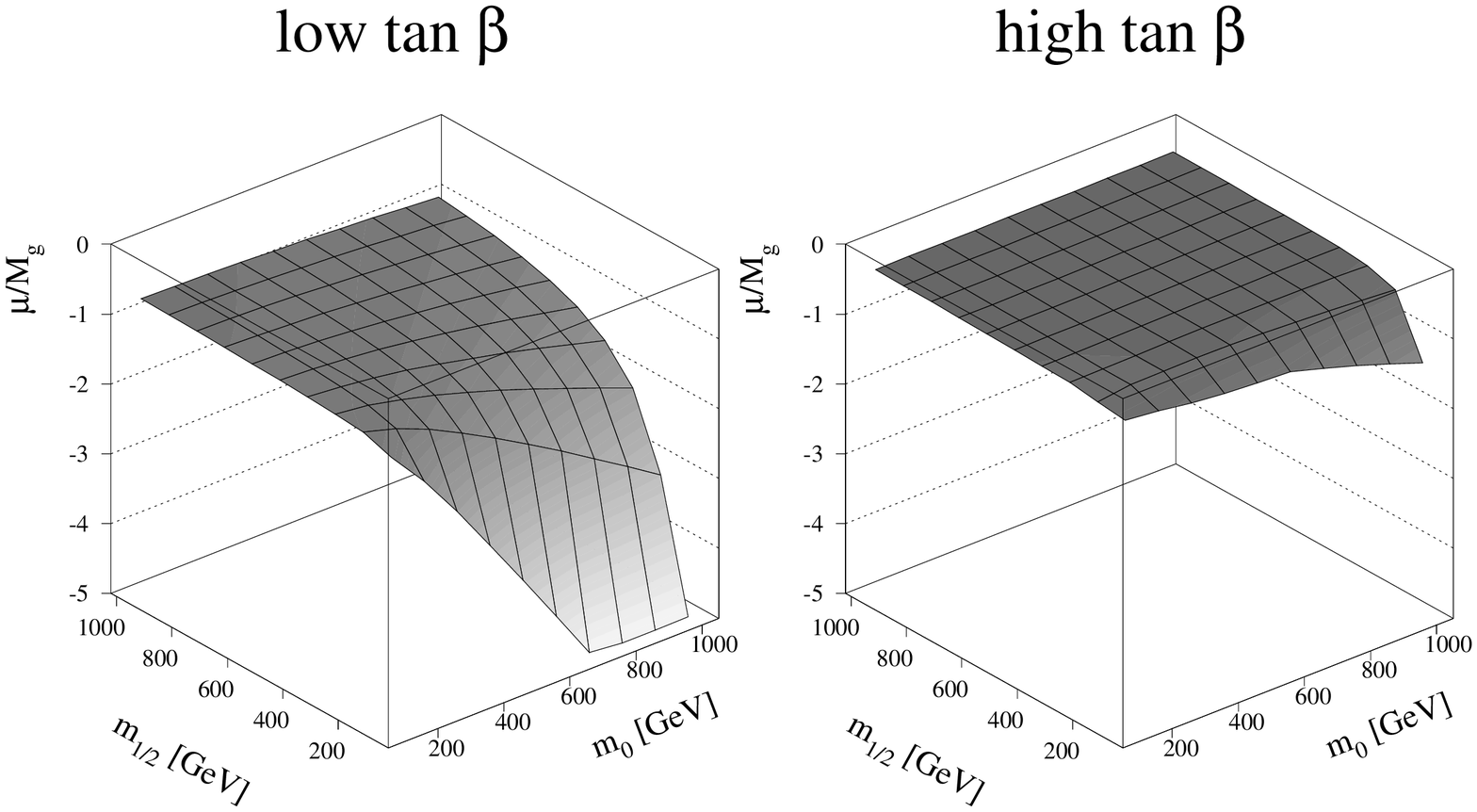}
\end{center}
\caption[]{\label{\figXII}The ratio of $\mu(\mz)$ and the gluino mass
               as function of
   $\mze$ and $\mha$   for the low and high $\tb$ scenario, respectively.}
\end{figure}

\begin{figure}
\begin{center}
  \leavevmode
  \epsfxsize=15cm
  \epsffile{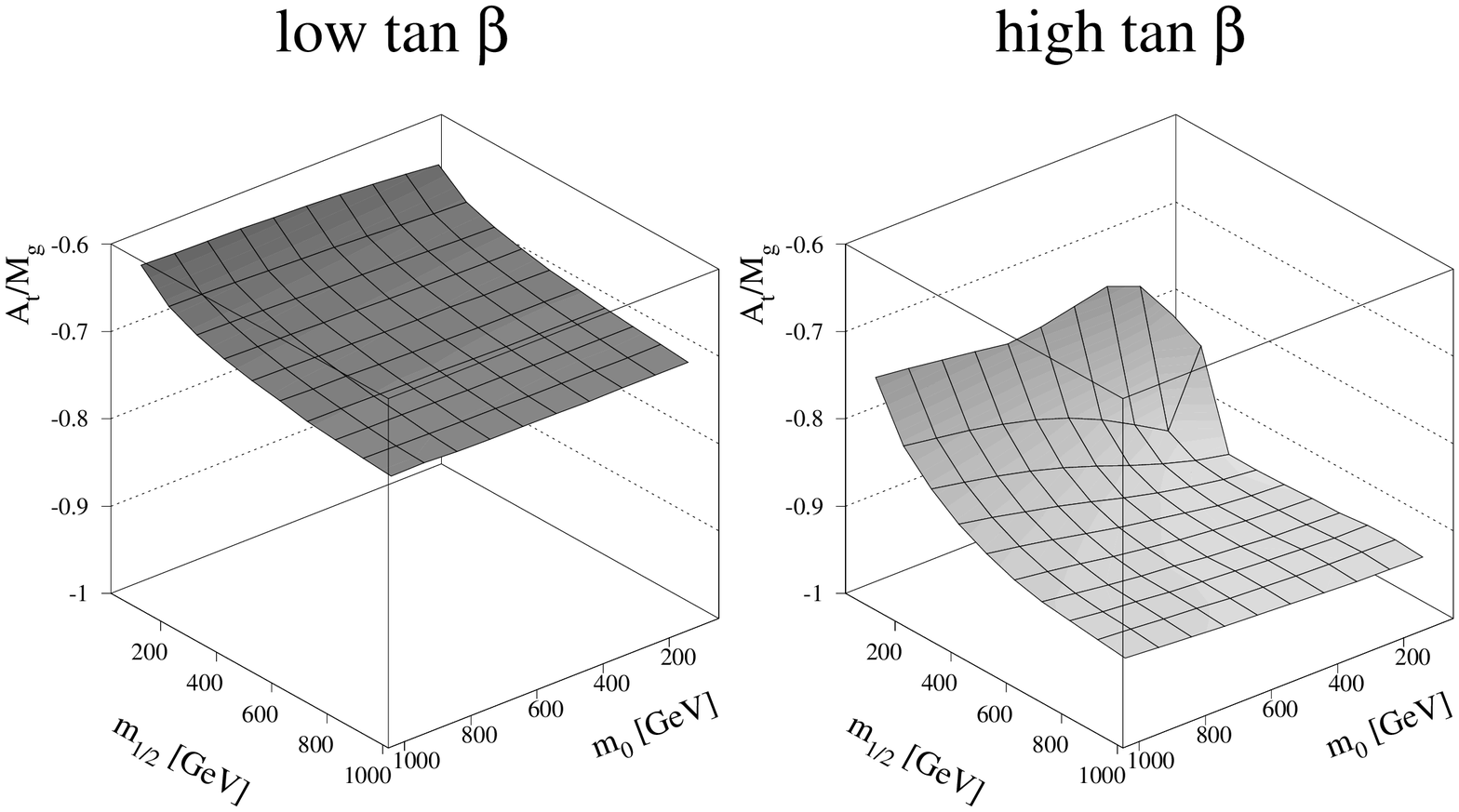}
\end{center}
\caption[]{\label{\figXIII}The ratio of $A_t(\mz)$ and the gluino mass as
function of
$\mze$ and $\mha$ for the low and high $\tb$ scenario, respectively.}
\end{figure}

\begin{figure}
\begin{center}
  \leavevmode
  \epsfxsize=15cm
  \epsffile{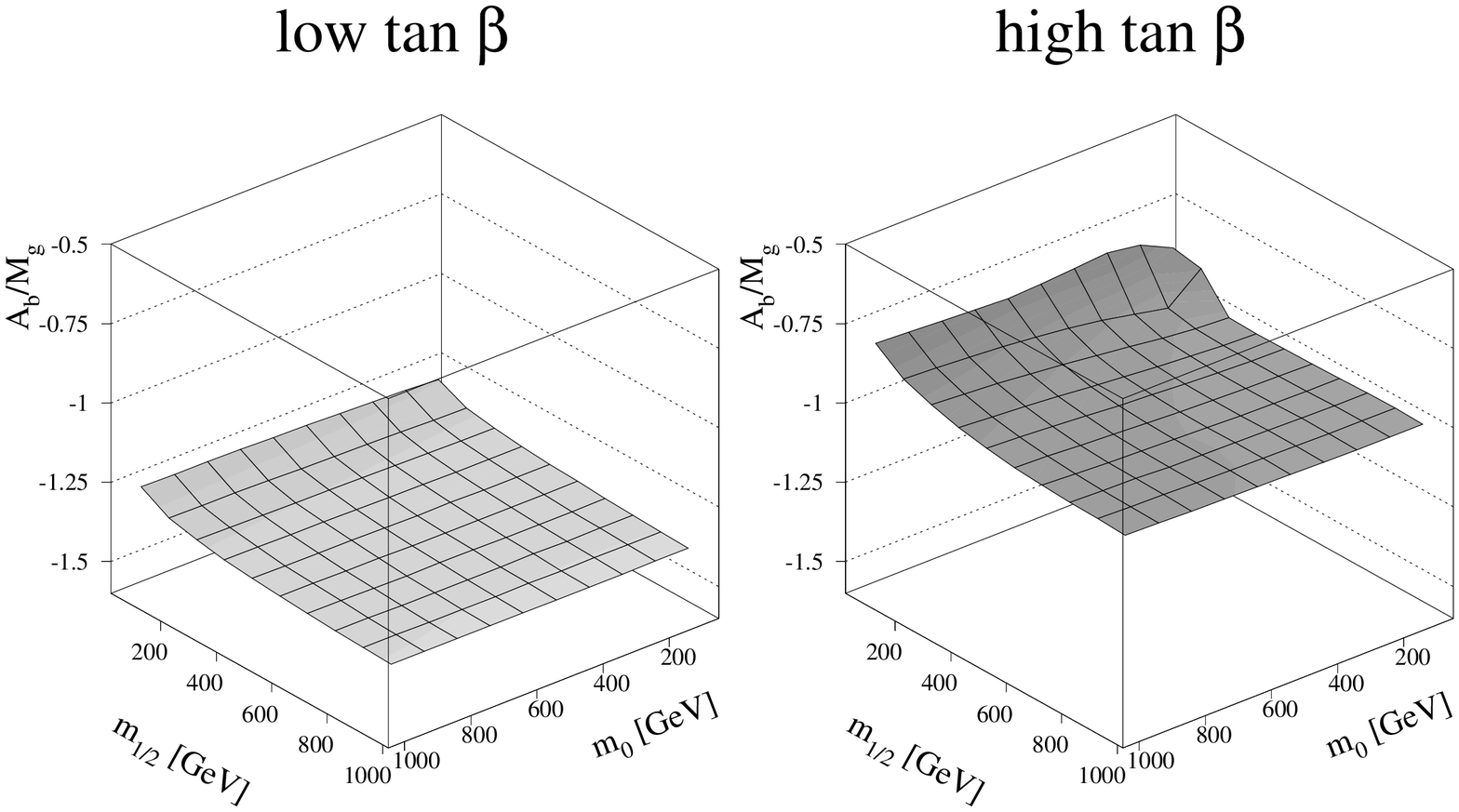}
\end{center}
\caption[]{\label{\figXIV}The ratio of $A_b(\mz)$ and the gluino mass as
function of
$\mze$ and $\mha$  for the low and high $\tb$ scenario, respectively.}
\end{figure}

\begin{figure}
\begin{center}
  \leavevmode
  \epsfxsize=15cm
  \epsffile{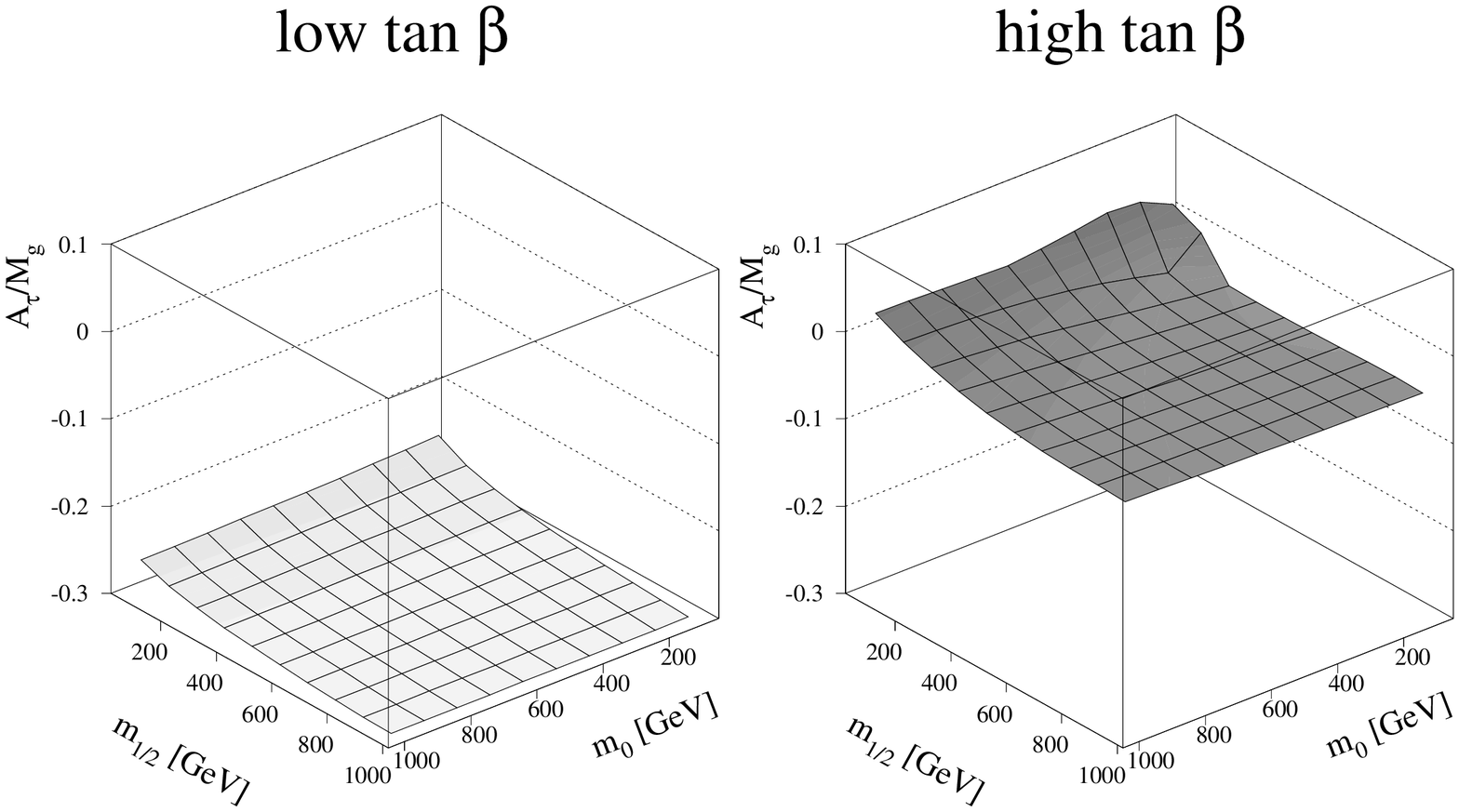}
\end{center}
\caption[]{\label{\figXV}The ratio of $A_\tau(\mz)$ and the gluino mass as
function of
$\mze$ and $\mha$  for the low and high $\tb$ scenario, respectively.}
\end{figure}

The fitted values of the trilinear couplings and the Higgs mixing
parameter $\mu$ are strongly correlated with $\mha$, so the ratio of these
parameters at the electroweak scale and the gluino mass, which is about
2.7~$\mha$ as shown in fig. \ref{\figXIII}, is relatively constant and
largely independent of $\mze$ (see  figs. \ref{\figXII} - \ref{\figXV}).
Note from the figures that although the trilinear couplings $A_t$, $A_b$
and $A_\tau$ have equal values at the GUT scale, they are quite
different at the electroweak scale due to the different RGE's.

\begin{figure}
\begin{center}
  \leavevmode
  \epsfxsize=15cm
  \epsffile{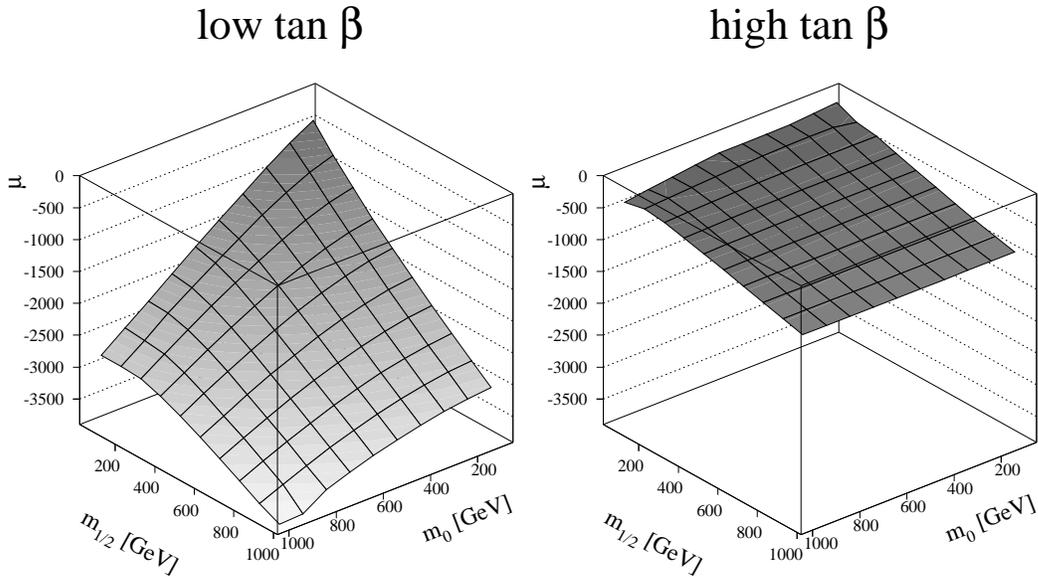}
\end{center}
\caption[]{\label{\figXVI}The Higgs mixing parameter $\mu_0$ at the GUT scale
          as function of $\mze$ and $\mha$
          for the low and high $\tb$ scenario, respectively.}
\end{figure}

The value of
$\mu$ at the GUT scale is shown in fig. \ref{\figXVI}. Note the large values
of $\mu$ at low $\tb$, which implies little mixing in the neutralino
sector and leads to eigenvalues of approximately $M_1$, $M_2$ and $\mu$
in the mass matrix (eq. \ref{neutmix}). Since $M_1$ is the smallest
value, the LSP will be almost purely a bino, which leads
to strong constraints on the parameters from the lifetime of the universe.

\begin{figure}
\begin{center}
  \leavevmode
  \epsfxsize=15cm
  \epsffile{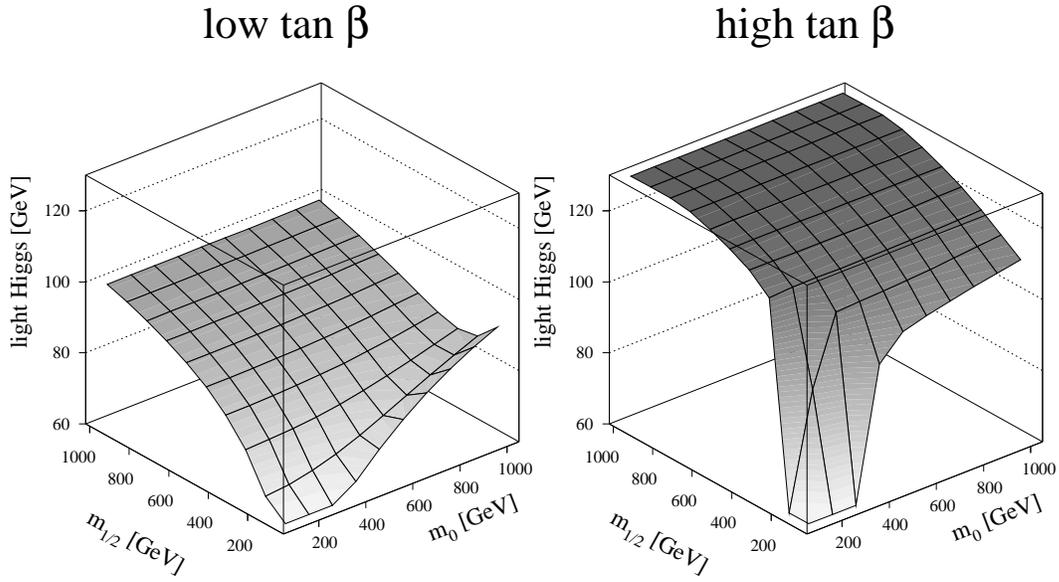}
\end{center}
\caption[]{\label{\figXVII}The mass of the lightest CP-even Higgs as function
    of $\mze$ and $\mha$ for the low and high $\tb$ scenario, respectively.
         The sign of $\mu $ is negative, as required for the high
           $\tb$ solution, but chosen negative for low $\tb$.
          }
\end{figure}

\begin{figure}
\begin{center}
  \leavevmode
  \epsfxsize=15cm
  \epsffile{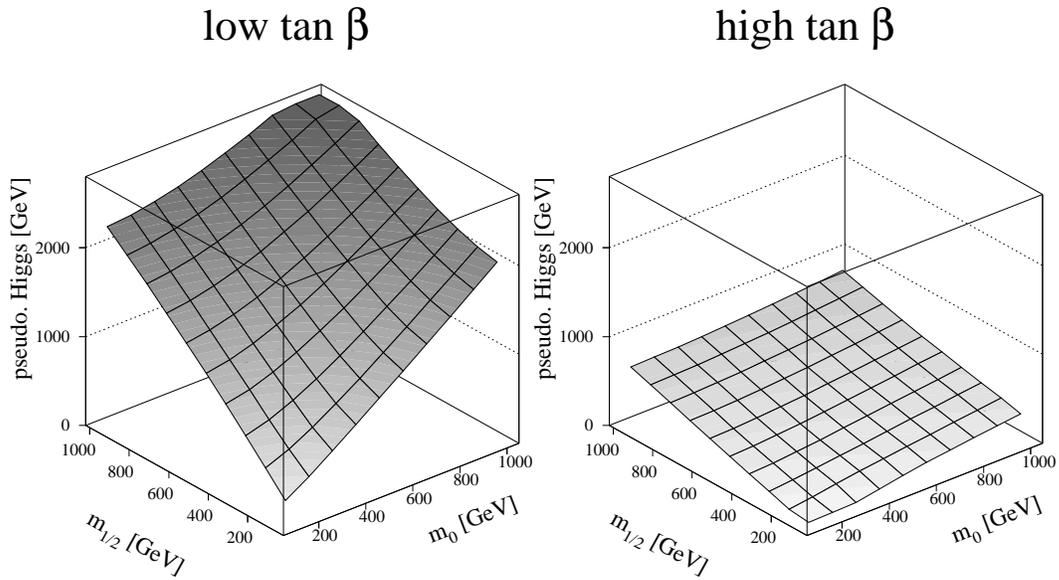}
\end{center}
\caption[]{\label{\figXVIII}The mass of the CP-odd Higgs as function of
         $\mze$ and $\mha$  for the low and high $\tb$ scenario, respectively.}
\end{figure}

Table \ref{t2a} shows the parameters from the best fit and table
\ref{t2} displays the corresponding SUSY masses. In figs.
\ref{\figXVII},\ref{\figXVIII} the masses of the lightest CP-even
and CP-odd Higgs bosons are shown for the whole parameter space
for negative $\mu$-values.
At each point a fit was performed to obtain the best solution
for the GUT parameters.  The mass of the lightest Higgs  saturates
at 100 GeV. For positive $\mu$-values and low $\tb$ the maximum Higgs mass
 increases to 115 GeV.

For high $\tb$ only negative $\mu$-
values are allowed, since positive $\mu$-values yield a too high b-mass
due to the large positive corrections in that case.
The upper limit on the Higgs mass for positive $\mu$ and {\it large} $\tb$
is about 130 GeV.

\begin{figure}
\begin{center}
  \leavevmode
  \epsfxsize=15cm
  \epsffile{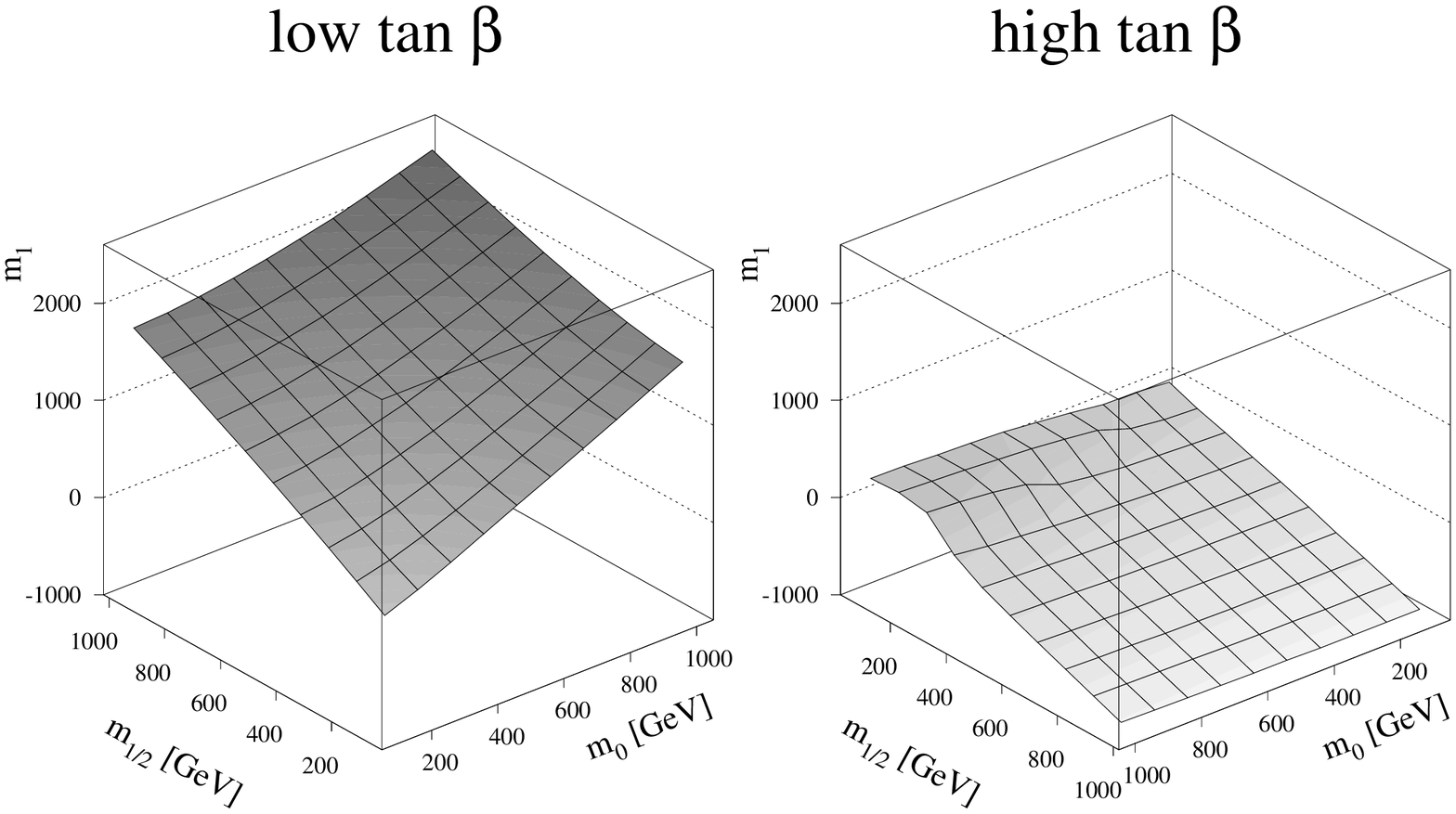}
\end{center}
\caption[]{\label{\figXIXa}The mass   $m_1$ in the Higgs potential
         at $\mz$ (Born level) in GeV as function of $\mze$ and $\mha$
          for the low and high $\tb$ scenario, respectively.
          }
\end{figure}

\begin{figure}
\begin{center}
  \leavevmode
  \epsfxsize=15cm
  \epsffile{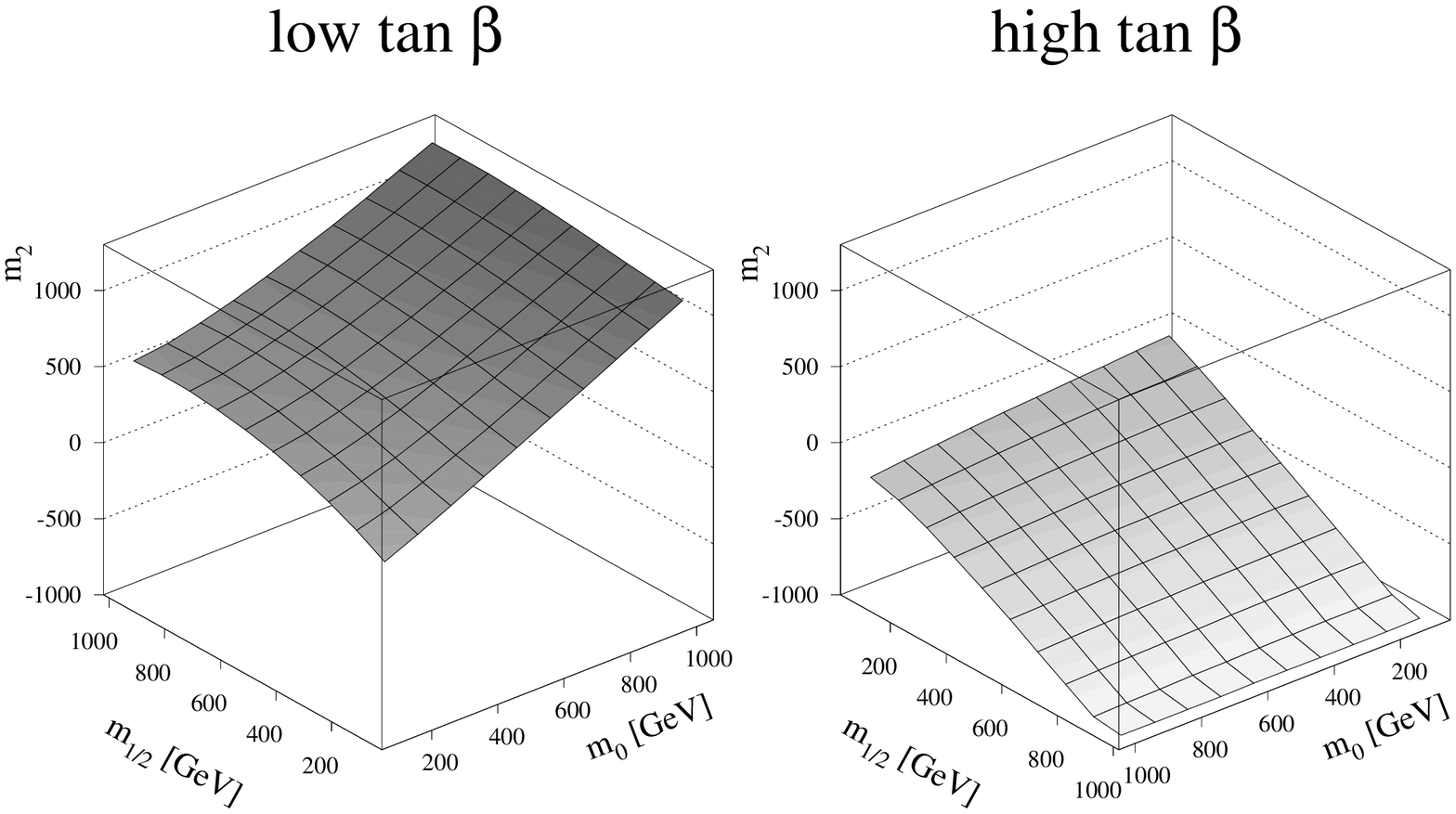}
\end{center}
\caption[]{\label{\figXIXb}The mass   $m_2$ in the Higgs potential
         at $\mz$ (Born level) in GeV as function of $\mze$ and $\mha$
         for the low and high $\tb$ scenario, respectively.
         }
\end{figure}

\begin{figure}
\begin{center}
  \leavevmode
  \epsfxsize=15cm
  \epsffile{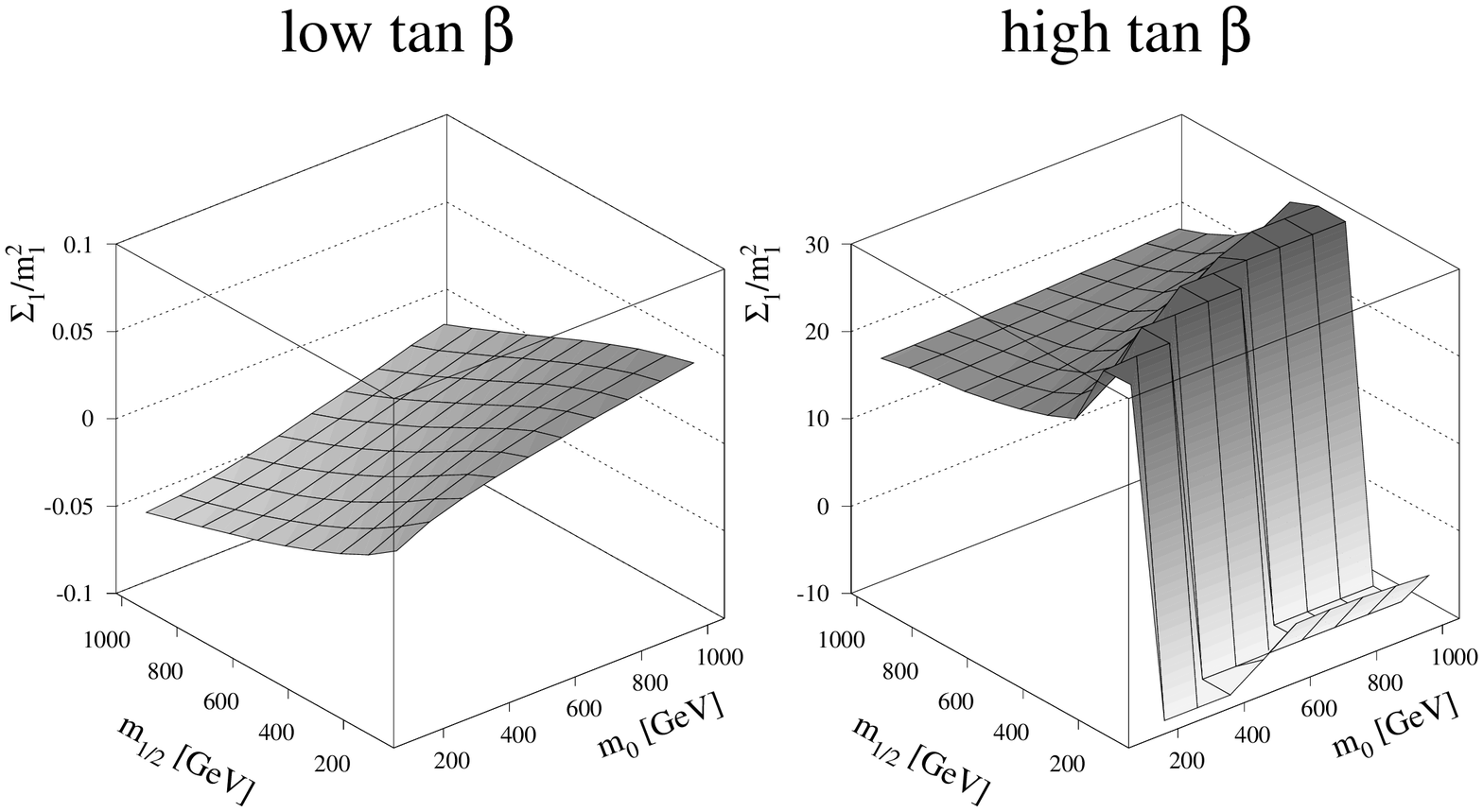}
\end{center}
\caption[]{\label{\figXIXc}The one-loop corrections  $\Sigma_1/m_1^2$
         at $\mz$
         as function of $\mze$ and $\mha$
          for the low and high $\tb$ scenario, respectively.}
\end{figure}

\begin{figure}
\begin{center}
  \leavevmode
  \epsfxsize=15cm
  \epsffile{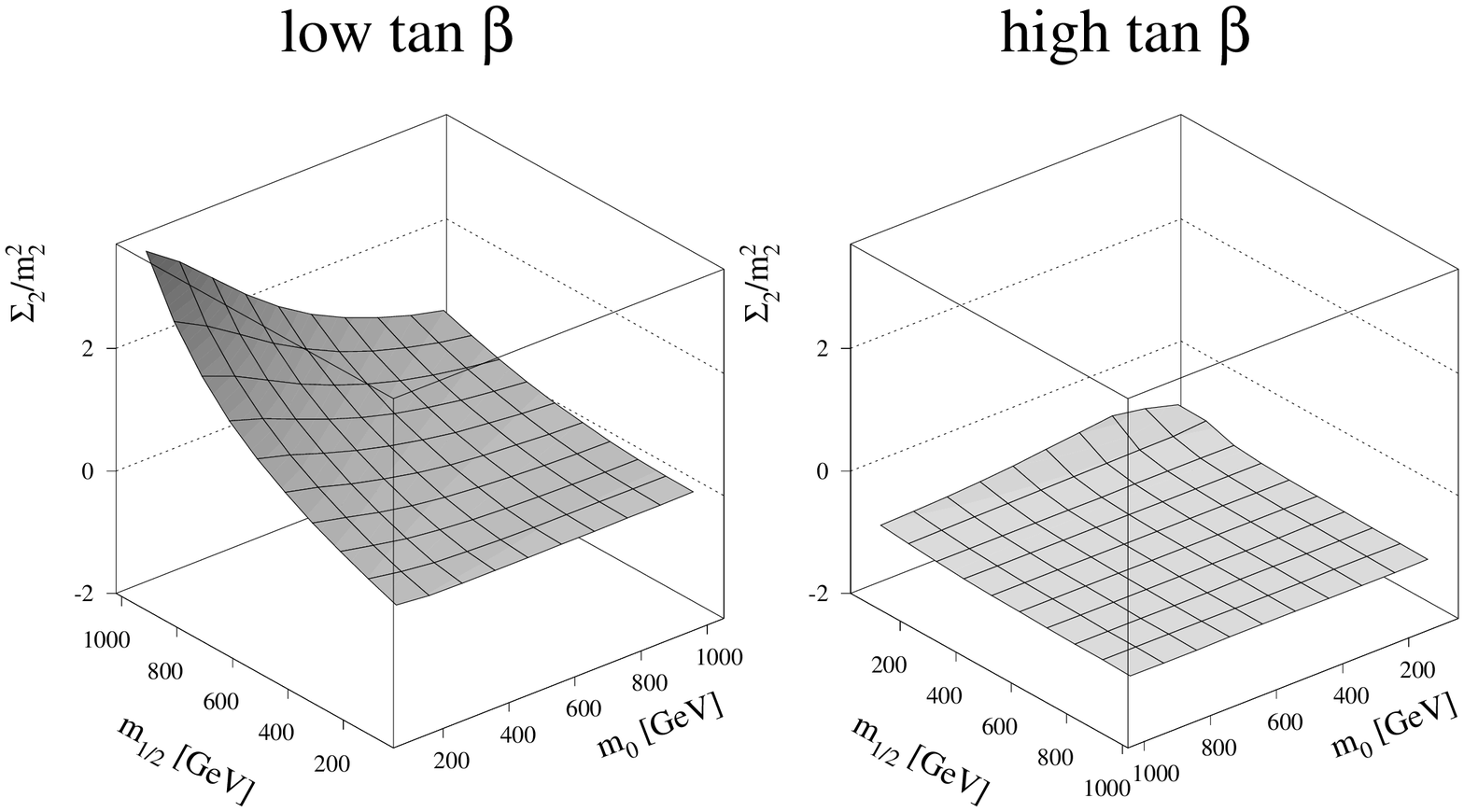}
\end{center}
\caption[]{\label{\figXIXd}The one-loop corrections  $\Sigma_2/m_2^2$
          at $\mz$
         as function of $\mze$ and $\mha$
         for the low and high $\tb$ scenario, respectively.}
\end{figure}

\begin{figure}
\begin{center}
  \leavevmode
  \epsfxsize=15cm
  \epsffile{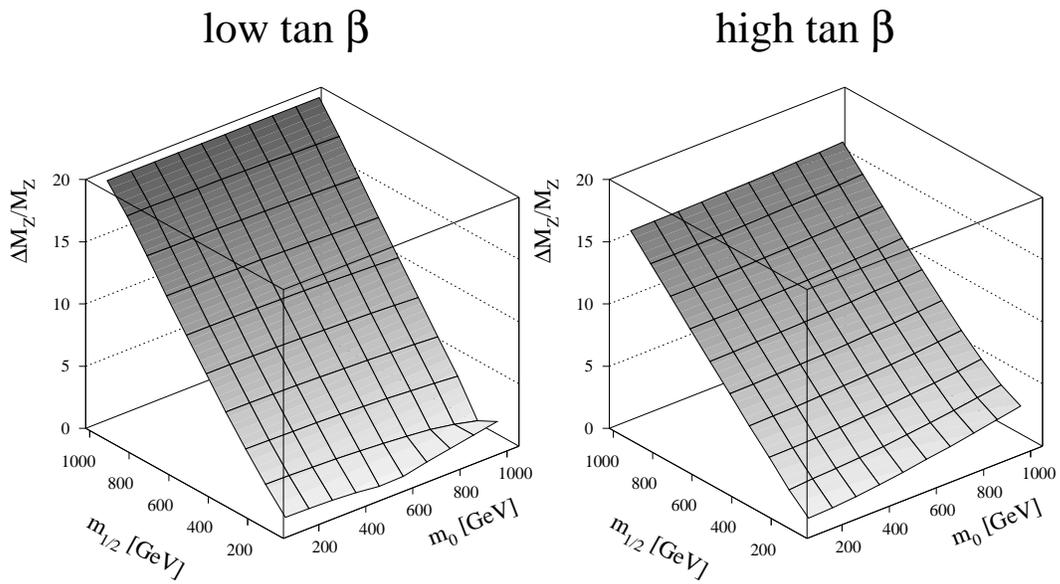}
\end{center}
\caption[]{\label{\figXIXe}The one-loop radiative corrections $\Delta(\mz)/\mz$
         as function of $\mze$ and $\mha$
         for the low and high $\tb$ scenario, respectively.}
\end{figure}

The upper limits for  particles other than the lightest Higgs are considerable
higher, unless restricted by some fine tuning argument: if the masses
of the superpartners and the normal particles are different, the famous
cancellation of quadratic divergencies in supersymmetry does not work
anymore and the corrections to the Higgs masses quickly increase, as shown
in figs. \ref{\figXIXa}-\ref{\figXIXd}.  These corrections lead to
large corrections in the electroweak scale too (see fig. \ref{\figXIXe}).
 It is a question of taste,
if one considers the corrections large or small and if one should exclude
some region of parameter space. In our opinion the fine tuning argument
is difficult to use for a mass scale below 1 TeV, and the whole region
up to 1 TeV should be considered, leading to quite
large upper limits~in case of the low $\tb$ scenario\cite{bek}.
\section{Discovery Potential  at LEP II}
The programs SUSYGEN~\cite{susygen}  and special SUSY
routines \cite{isajetee} in
ISAJET~\cite{isajet} have been used to calculate the  production cross-sections
for charginos, neutralinos and the lightest Higgses as function of the
SUSY mass scales $\mze$ and $\mha$.

\begin{figure}
\begin{center}
  \leavevmode
  \epsfxsize=15cm
  \epsffile{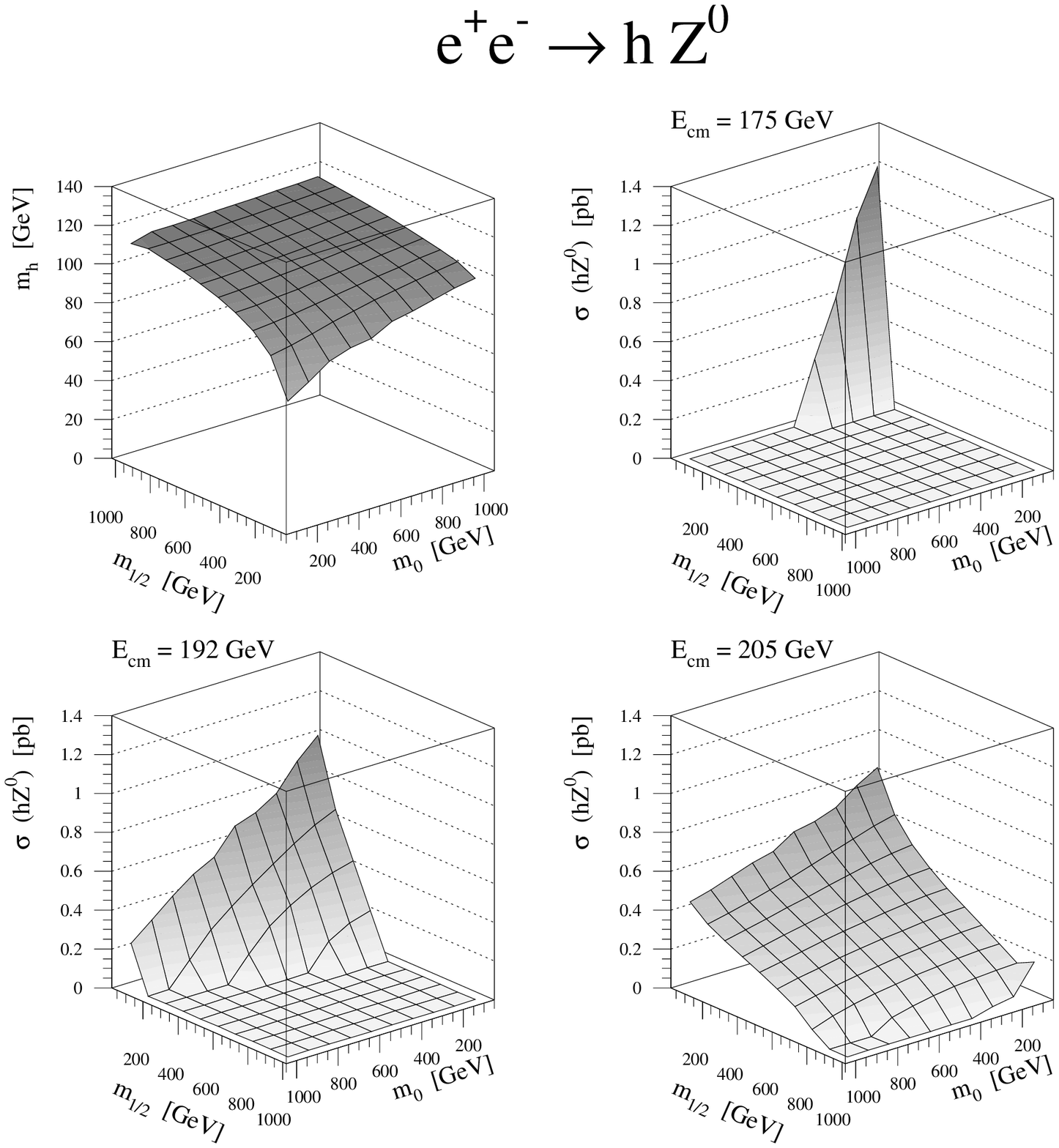}
\end{center}
\caption[]{\label{\figXX}The Higgs mass as function of $\mze$ and $\mha$ for
positive values of $\mu$ and low $\tb$  (left top corner) and the main
production cross sections for three different LEP II energies (175, 192 and
205 GeV). For negative $\mu$ values the Higgs mass  is lighter
(see fig. \ref{\figVII}) and the cross sections   about 50\% larger.
}
\end{figure}

Fig. \ref{\figXX} shows the mass of the lightest Higgs boson and the
corresponding Higgs production cross sections at three LEP  energies
as functions of $m_0$ and $m_{1/2}$ for $\tan\beta=1.7$.
The absolute value of the Higgs mass parameter $\mu$
was determined from the electroweak symmetry breaking condition;
its sign was chosen positive. For negative $\mu$ values the
cross sections are about 50\% higher due to the lighter
Higgs mass in that case (see fig. \ref{\figXVII}). Note the
strong dependence of the cross section on the LEP centre-of-mass
energy. At 205~GeV the whole CMSSM parameter space is covered,
since at even higher values of the SUSY scale the Higgs mass
hardly increases, as shown in the left top corner of fig. \ref{\figXX}.
For these results the large radiative corrections to $A_t$ and $A_b$
were taken into account, so they have nonzero values at the electroweak
scale, typically $A_t = -0.7M_{\tilde{g}}$ and $A_b = -1.5M_{\tilde{g}}$
(see figs. \ref{\figXII} - \ref{\figXV}). Note the fixed point
behaviour of $A_t$ at low $\tb$ values, as shown before in  fig. \ref{\figXa}.

\begin{figure}
\begin{center}
  \leavevmode
  \epsfxsize=15cm
  \epsffile{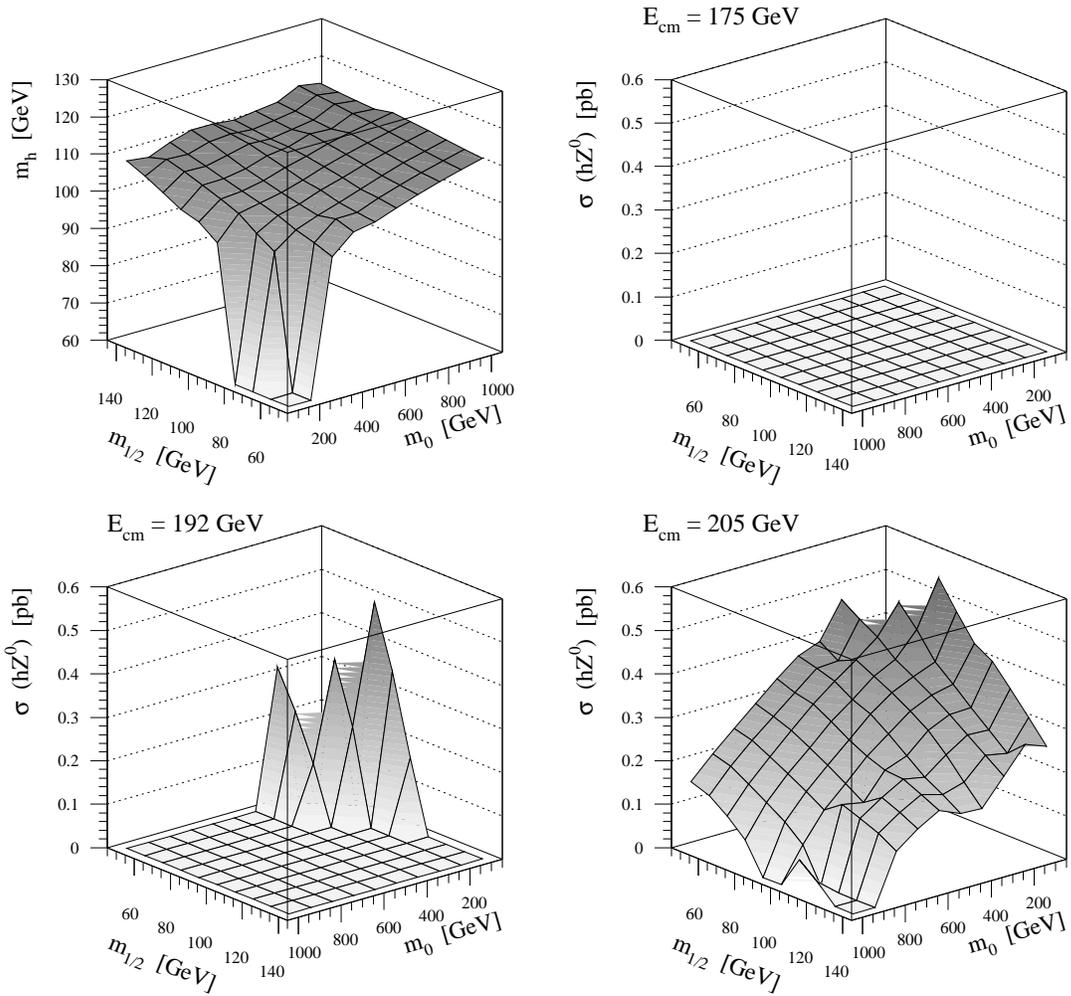}
\end{center}
\caption[]{\label{\figXXI}The Higgs mass as function of $\mze$ and $\mha$ for
$\mu_0 = -300$ GeV for   $\tb=46$    (left top corner) and the main
production cross sections for three different LEP II energies (175, 192 and
205 GeV).
         $\mu$ was kept
         to a representative value     (see fig. \ref{\figXVI}),
         since in most of
         the region the fit gave an
         unacceptable $\chi^2$, so $\mu$ could not be determined.
  }
\end{figure}

\begin{figure}
\begin{center}
  \leavevmode
  \epsfxsize=15cm
  \epsffile{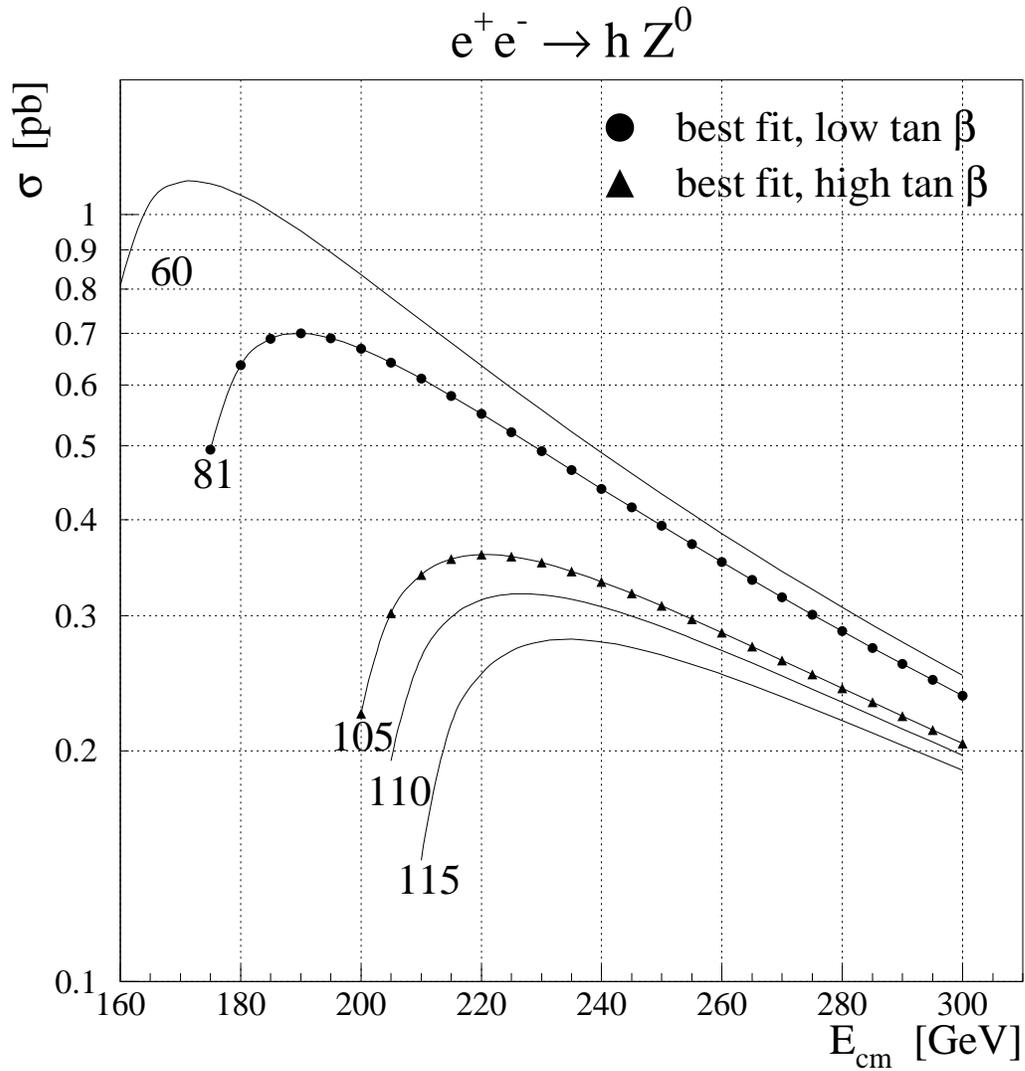}
\end{center}
\caption[]{\label{\figXXII}
         The cross section as function of the center of mass
         energy for different Higgs masses, as indicated by the numbers (in
         GeV). The upper limit on the Higgs mass
         is $\approx$ 115 GeV.  (see figs. \ref{\figXX} and \ref{\figXXI}).
         }
\end{figure}

\begin{figure}
\begin{center}
  \leavevmode
  \epsfxsize=15cm
  \epsffile{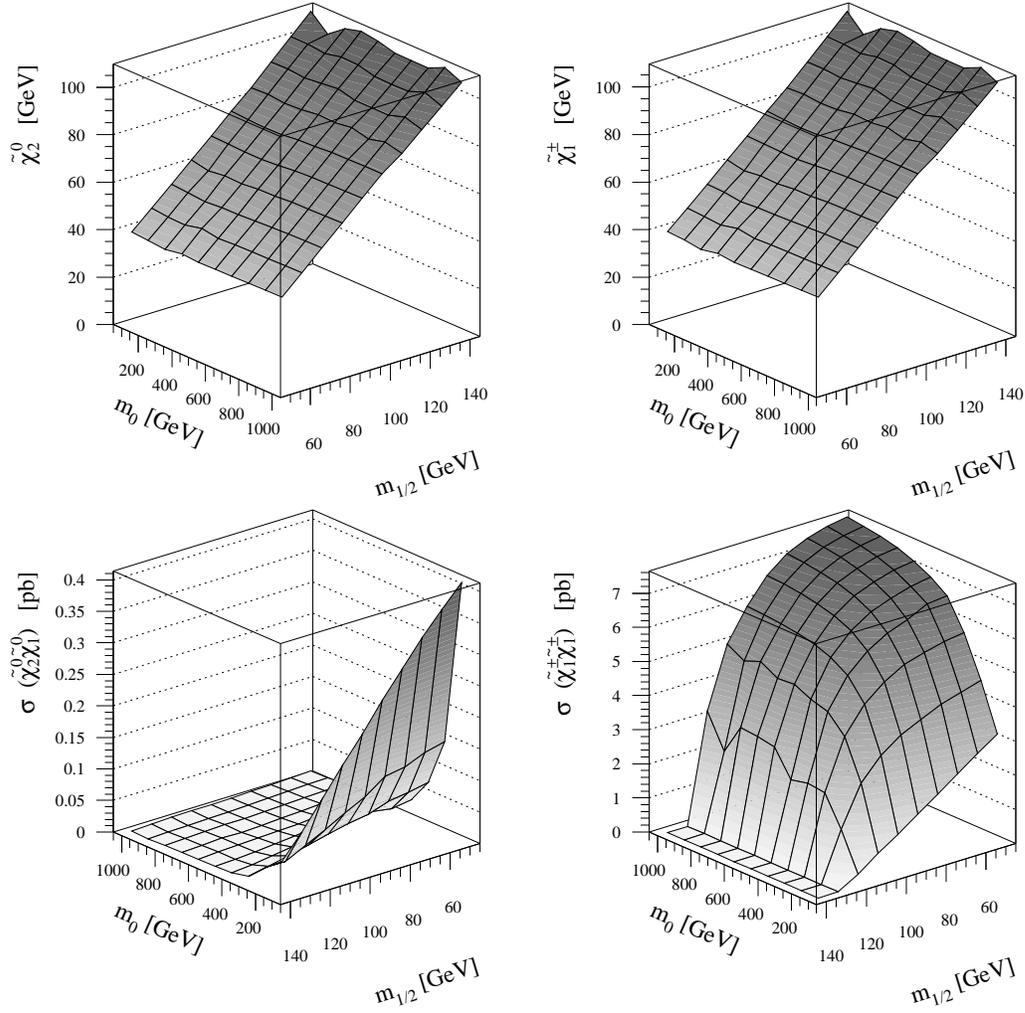}
\end{center}
\caption[]{\label{\figXXIII}The masses of the second lightest  neutralino
         and chargino masses as well as the production
         cross sections for  $\mu_0=-300$ GeV and $\tb = 46$. $\mu$ was kept
         to a representative value (see fig. \ref{\figXVI}), since in most
         of the region the fit gave an
         unacceptable $\chi^2$, so $\mu$ could not be determined.
         Positive values of $\mu$ give similar results.
         The steep decrease in the chargino cross section at small values
         of $\mze$ is due to the light sneutrino in that region, which
         leads to a strong negative interference between s- and t-channel.
           Fortunately, the neutralino production is large there,
           as shown by the plot in the left bottom corner.
         }
\end{figure}

For large values of $\tb$  only negative values of $\mu$ give acceptable
fits, mainly because of the large corrections to the bottom mass, which
prohibit tau-bottom unification for positive $\mu$. In this case
the lightest Higgs mass becomes
as large as 130 GeV for    large values of the SUSY scales $\mze$ and $\mha$.
However, if one includes the constraint from the radiative
$b\rightarrow s\gamma$ decay, as measured by CLEO~\cite{cleo94},  only
low values of $m_{1/2}$ give acceptable fits, in which case the
remaining parameter space is largely accessible to a Higgs search,
provided LEP II reaches its highest energies (see Fig.~\ref{\figXXI}).
The Higgs cross section is mainly a function of the Higgs mass and the center
of mass energy. This dependency is shown in fig.~\ref{\figXXII} for some
representative Higgs masses. In addition, the chargino
and neutralino searches cover these regions, as shown in Fig.~\ref{\figXXIII}:
even at a LEP II energy of 192 GeV
searches   for both, neutralino and chargino production, cover
the region $\mha < 110$ GeV, which is the
region of interest for large \tb~(see fig.~\ref{\figIII}).

\section{Summary}
In the Constrained Minimal Supersymmetric Model (CMSSM)
the optimum values of the GUT scale parameters and the
corresponding SUSY mass spectra for the low
and high $\tb$ scenario have been determined from a combined fit
 to the low energy data on couplings, quark and lepton masses of the third
generation, the electroweak scale $\mz$,
  \bsg, and the lifetime of the universe.

At the highest LEP II energy of 205 GeV practically the whole parameter
space of the CMSSM can be covered, both for the low and high $\tb$
scenario, if one searches for Higgses, charginos and neutralinos.
 At the lower  envisaged LEP II energy of 192 GeV
only half of the parameter space for low $\tb$  is accessible.
\section{Acknowledgment}
We thank Drs. D.I. Kazakov and  A.V. Gladyshev for useful discussions.
The research described in this publication was made possible
in part by support from the Human Capital and Mobility Fund
(Contract ERBCHRXCT 930345) from the European Community, and by support
from the German Bundesministerium f\"ur Bildung und Forschung (BMBF)
(Contract 05-6KA16P).
\clearpage
%
%--------------- mt_tb 0
%
%
%--------------- m_1/m_2 vs.Q
%
%
%--------------- chi^2
%
%
%--------------- spectrum
%
\begin{figure}
 \begin{center}
  \leavevmode
  \epsfxsize=15cm
  \epsffile{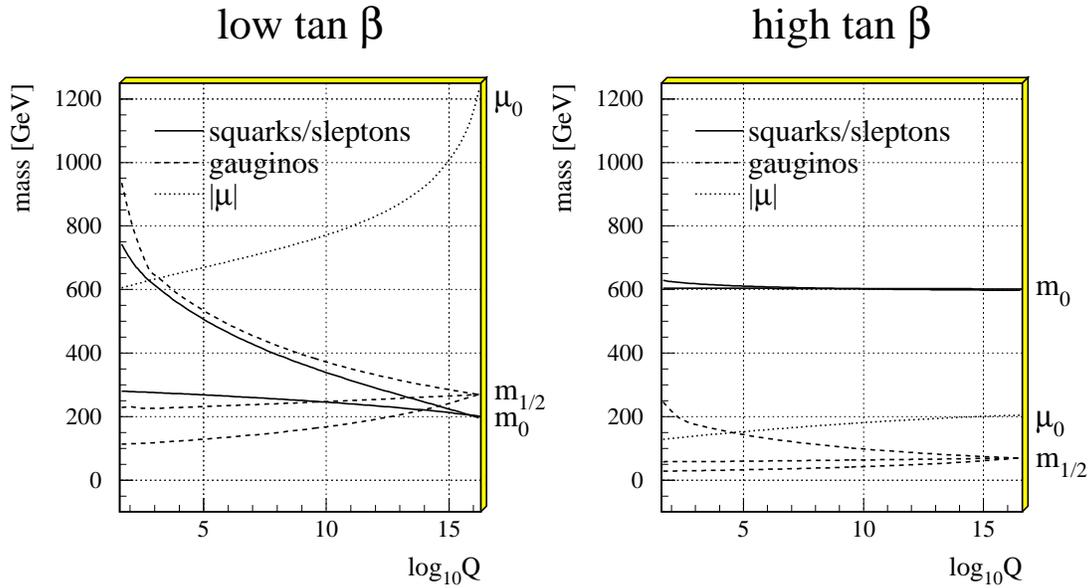}
\end{center}
\caption[]{\label{figIV}The running of the particle masses and the
                $\mu$ parameter for low and high
                $\tb$ values.}
\end{figure}
%
%--------------- b --> s gamma
%
%
%--------------- relic density
%
%
%--------------- m_charg (new) fig VII
%
%
%--------------- chi^2
%
%
%--------------- Delta m_b
%
%
%--------------- mt_tb +/- Dmb, +/- bsg
%
%
%----------------- A_t running
%
%\clearpage
%
%----------------- m_gluino
%
\begin{figure}
\begin{center}
  \leavevmode
  \epsfxsize=15cm
  \epsffile{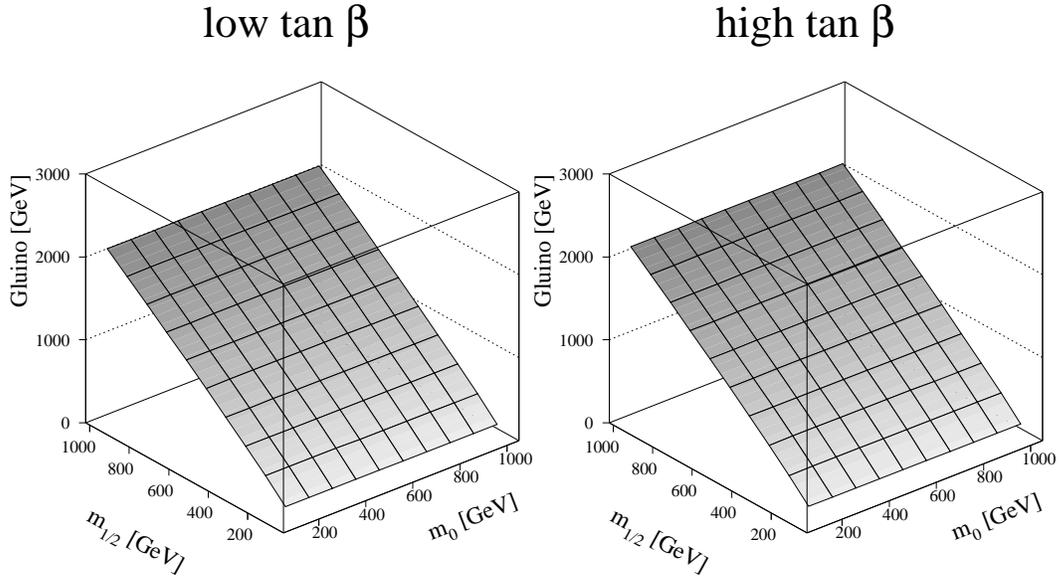}
\end{center}
\caption[]{\label{\figXI}The gluino mass as function of $\mze$ and $\mha$
                for the low and high $\tb$ scenario, respectively.}
\end{figure}
%
%----------------- mu/m_gluino
%
%
%----------------- A_top/m_gluino
%
%
%----------------- A_bot/m_gluino
%
%
%----------------- A_tau/m_gluino
%
\clearpage
%
%----------------- \mu
%
%
%----------------- mh
%
%
%----------------- ma
%
%
%----------------- m1
%
%
%----------------- m2
%
%
%----------------- sigma1/m1
%
%
%----------------- sigma2/m2
%
%
%----------------- dmz
%
%
%----------------- 43a_hz0 low cross sections
%

%
%----------------- 49a_hz0-300 high cross sections
%
%
%-----------------  x_hz0_low_high cross sections
%

%
%----------------- 49a_chichi-300 cross sections
%
\clearpage

%----------------------------------------------------------------------------
\clearpage
%\addcontentsline{toc}{chapter}{References.}
%   \bibliographystyle{unsrt}
%   \bibliography{biblio}

\begin{thebibliography}{10}

\bibitem{su5susy}
{\em {\rm P.~Fayet}, Phys. Lett. {\bf B64} (1976) 159; ibid. {\bf B60} (1977)
  489; \\ {\rm S.~Dimopoulos, H.~Georgi}, Nucl. Phys. {\bf B193} (1981) 150; \\
  {\rm L. E. Ib\'a\~nez, G. G. Ross}, Phys. Lett. {\bf B105} (1981) 435; \\
  {\rm S. Dimopoulos, S. Raby, F. Wilczek}, Phys. Rev. {\bf D24} (1981) 1681;\\
  {\rm N.~Sakai}, Z.~Phys.~{\bf C11} (1981) 153;\\ {\rm A.~H.~Chamseddine,
  R.~Arnowitt and P.~Nath}, Phys.~Rev.~Lett.~{\bf 49} (1982) 970.}

\bibitem{ekn2}
J.~Ellis, S.~Kelley, and D.~V. Nanopoulos.
\newblock {\em Nucl.~Phys.~{\bf B373} (1992) 55.}

\bibitem{abf}
U.~Amaldi, W.~de~Boer, and H.~F\"urstenau.
\newblock {\em Phys.~Lett.~{\bf B260} (1991) 447.}

\bibitem{lalu}
P.~Langacker and M.~Luo.
\newblock {\em Phys.~Rev.~{\bf D44} (1991) 817.}

\bibitem{rev}
{\em {\rm For references see the review papers: \\ H.-P. Nilles, Phys. Rep.
  {\bf 110} (1984) 1;\\ H.E. Haber, G.L. Kane, Phys. Rep. {\bf 117} (1985)
  75;\\ A.B. Lahanas and D.V. Nanopoulos, Phys. Rep. {\bf 145} (1987) 1; \\ R.
  Barbieri, Riv. Nuo. Cim. {\bf 11} (1988) 1.}}

\bibitem{rrb92}
G.~G. Ross and R.~G. Roberts.
\newblock {\em Nucl.~Phys.~{\bf B377} (1992) 571.}

\bibitem{cpw}
M.~Carena, S.~Pokorski, and C.~E.~M. Wagner.
\newblock {\em Nucl.~Phys.~{\bf B406} (1993) 59.}

\bibitem{bbo}
V.~Barger, M.~S. Berger, and P.~Ohmann.
\newblock {\em Phys.~Rev.~{\bf D47} (1993) 1093.}

\bibitem{op}
M.~Olechowsi and S.~Pokorski.
\newblock {\em Nucl.~Phys.~{\bf B404} (1993) 590.}

\bibitem{chan}
P.H.~Chankowski et~al.
\newblock {\em IFT-95-9, MPI-PTH-95-58 and ref. therein.}

\bibitem{langpol}
P.~Langacker and N.~Polonsky.
\newblock {\em Phys.~Rev.~{\bf D49} (1994) 1454; UPR-0642-T and ref. therein.}

\bibitem{zic}
J.~L. Lopez, D.V. Nanopoulos, and A.~Zichichi.
\newblock {\em Progr. in Nucl. and Particle Phys., {\bf 33} (1994) 303 and ref.
  therein.}

\bibitem{wdb}
W.~de~Boer.
\newblock {\em Progr. in Nucl. and Particle Phys., {\bf 33} (1994) 201}.

\bibitem{bek}
W.~de~Boer, R.~Ehret, and D.~Kazakov.
\newblock {\em Phys. Lett. {\bf B334} (1994) 220}.

\bibitem{nanopo}
{\em J. L. Lopez, D.V. Nanopoulos, A. Zichichi, CTP-TAMU-40-93 (1993);
  CTP-TAMU-33-93 (1993); CERN-TH-6934-93 (1993); CERN-TH-6926-93-REV (1993);
  CERN-TH-6903-93 (1993);\\ J. L. Lopez, et al., Phys. Lett. {\bf B306} (1993)
  73.}

\bibitem{ram1}
{\em {\em S.P. Martin and P. Ramond}, Phys. Rev. {\bf D48} (1993) 5365\\ {\em
  D.J. Castano, E.J. Piard, and P. Ramond,} Phys. Rev. {\bf D49} (1994) 4882}.

\bibitem{roskane}
G.~L. Kane, C.~Kolda, L.~Roszkowski, and J.~D. Wells.
\newblock {\em Phys.~Rev.~{\bf D49} (1994) 6173}.

\bibitem{roskane1}
C.~Kolda, L.~Roszkowski, and J.~D.~Wells and G.~L. Kane.
\newblock {\em Phys.~Rev.~{\bf D50} (1994) 3498}.

\bibitem{cawa}
M.~Carena and C.~E.~M. Wagner.
\newblock {\em CERN-TH-7393-94 and ref. therein.}

\bibitem{copw}
M.~Carena, M.~Olechowski, S.~Pokorski, and C.~E.~M. Wagner.
\newblock {\em Nucl.~Phys.~{\bf 419} (1994) 213.}

\bibitem{cleo94}
R.~Ammar et~al. CLEO-Collaboration.
\newblock {\em Phys. Rev. Lett. {\bf 74} (1995) 2885}.

\bibitem{arn}
R.~Arnowitt and P.~Nath.
\newblock {\em Phys.~Rev.~Lett.~{\bf 69} (1992) 725; Phys.~Lett.~{\bf B287}
  (1992) 89; Phys.~Lett.~{\bf B289} (1992) 368; Phys.~Lett.~{\bf B299} (1993)
  58, ERRATUM-ibid.~{\bf B307} (1993) 403; Phys.~Rev.~Lett.~{\bf 70} (1993)
  3696; \\CTP-TAMU-23/93 (1993)}.
\newblock and references therein.

\bibitem{langac}
P.~Langacker.
\newblock {\em Univ.~of Penn.~Preprint, UPR-0539-T (1992)}.

\bibitem{flipsu5}
D.V.~Nanopoulos J.~Ellis, J.L.~Lopez.
\newblock {\em Phys.~Lett.~{\bf B252} (1990) 53.}

\bibitem{finite}
{D.I. Kazakov et al., }.
\newblock {\em Contr. paper to the EPS Conf., Brussels, (1995).}

\bibitem{susy}
{\em {\rm Yu.A.~Gol'fand and E.P.~Likhtman}, JETP Lett.~{\bf 13} (1971) 323; \\
  {\rm D.V.~Volkov and V.P.~Akulow}, Phys.~Lett.~{\bf 46b} (1971) 323; \\ {\rm
  J.~Wess and B.~Zumino}, Nucl.~Phys.~{\bf B70} (1974) 39; \\ {\rm J.~Wess and
  B.~Zumino}, Phys.~Lett.~{\bf 49B} (1974) 52; \\ {\rm S.~Ferrara, J.~Wess, and
  B.~Zumino.}, Phys.~Lett.~{\bf 51B} (1974) 239;\\ {\rm J.~Iliopoulos and
  B.~Zumino}, Nucl.~Phys.~{\bf B76} (1974) 310.}

\bibitem{loopewbr}
R.~Arnowitt and P.~Nath.
\newblock {\em Phys.~Rev.~{\bf D46} (1992) 3981.}

\bibitem{erz}
J.~Ellis, G.~Ridolfi, and F.~Zwirner.
\newblock {\em Phys.~Lett.~{\bf B257} (1991) 83; \\ Phys.~Lett.~{\bf B262}
  (1991) 477.}

\bibitem{berz}
A.~Brignole, J.~Ellis, G.~Ridolfi, and F.~Zwirner.
\newblock {\em Phys.~Lett.~{\bf B271} (1991) 123.}

\bibitem{drno}
M.~Drees and M.~M. Nojiri.
\newblock {\em Phys.~Rev.~{\bf D45} (1992) 2482.}

\bibitem{kz92}
Z.~Kunszt and F.~Zwirner.
\newblock {\em Nucl.~Phys.~{\bf B 385} (1992) 3.}

\bibitem{eqz}
J.~R. Espinosa, M.~Quir\'os, and F.~Zwirner.
\newblock {\em Phys.~Lett.~{\bf B307} (1993) 106.}

\bibitem{cpr}
P.~H. Chankowski, S.~Pokorski, and J.~Rosiek.
\newblock {\em Nucl. Phys. {\bf B423 (1994) 437}; \\MPI-PH-92-116 (1992);
  MPI-PH-92-116-ERR (1992).}

\bibitem{abfI}
U.~Amaldi, W.~de~Boer, P.~H. Frampton, H.~F\"urstenau, and J.T. Liu.
\newblock {\em \\Phys.~Lett.~{\bf B281} (1992) 374.}

\bibitem{yana}
{\em {\rm H.~Murayama and T.~Yanagida,} Mod. Phys. Lett. {\bf A7} (1992) 147;\\
  {\rm T.~G.~Rizzo}, Phys.~Rev.~{\bf D45} (1992) 3903;\\ {\rm T.~Moroi,
  H.~Murayama and T.~Yanagida}, Phys. Rev. {\bf D48} (1993) 2995.}

\bibitem{msbar}
{\em {\rm G.~'t Hooft}, Nucl.~Phys.~{\bf B61}, (1973) 455; \\ {\rm
  W.~A.~Bardeen, A.~Buras, D.~Duke and T.~Muta}, Phys.~Rev.~{\bf D 18}, (1978)
  3998.}

\bibitem{LEP}
{The LEP Collaborations: ALEPH, DELPHI, L3 and OPAL and the LEP electroweak
  Working Group;}.
\newblock {\em Phys.~Lett.~{\bf 276B} (1992) 247;}.
\newblock Updates are given in CERN/PPE/93-157, CERN/PPE/94-187 and
  LEPEWWG/95-01 (see also ALEPH 95-038, DELPHI 95-37, L3 Note 1736 and OPAL
  TN284.

\bibitem{PDB}
Review of~Particle~Properties.
\newblock {\em Phys.~Rev.~{\bf D50} (1994).}

\bibitem{CDF}
{CDF Collab.,~F.~Abe, et al.}
\newblock {\em Phys.~Rev.~Lett.~{\bf 74} (1995) 2626.}

\bibitem{D0}
{D0 Collab.,~S.~Abachi, et al.}
\newblock {\em Phys.~Rev.~Lett.~{\bf 74} (1995) 2632.}

\bibitem{dfs}
G.~Degrassi, S.~Fanchiotti, and A.~Sirlin.
\newblock {\em Nucl.~Phys.~{\bf B351} (1991) 49.}

\bibitem{EJ95}
S.~Eidelmann and F.~Jegerlehner.
\newblock {\em Hadronic contributions to (g-2) of leptons and to the effective
  fine structure constant $\alpha(m_Z^2)$, {\rm PSI Preprint PSI-PR-95-1.}}

\bibitem{akt}
I.~Antoniadis, C.~Kounnas, and K.~Tamvakis.
\newblock {\em Phys.~Lett.~{\bf 119B} (1982) 377.}

\bibitem{Ross1}
B.~Pendleton and G.G. Ross.
\newblock {\em Phys.~Lett.~{\bf B98} (1981) 291;}.

\bibitem{bbog}
V.~Barger, M.~S. Berger, P.~Ohmann, and R.~J.~N. Phillips.
\newblock {\em Phys.~Lett.~{\bf B314} (1993) 351.}

\bibitem{lanpol}
P.~Langacker and N.~Polonski.
\newblock {\em Univ.~of Pennsylvania Preprint UPR-0556-T, (1993).}

\bibitem{bmaskln}
S.~Kelley, J.~L. Lopez, and D.V. Nanopoulos.
\newblock {\em Phys.~Lett.~{\bf B274} (1992) 387.}

\bibitem{copw1}
M.~Carena, , M.~Olechowski, S.~Pokorski, and C.~E.~M. Wagner.
\newblock {\em Nucl.~Phys.~{\bf B426} (1994) 269.}

\bibitem{ara91}
H.~Arason et~al.
\newblock {\em Phys.~Rev.~Lett.~{\bf 67} (1991) 2933.}

\bibitem{runmas}
{\em {\rm J.~Gasser and H.~Leutwyler}, Phys.~Rep.~{\bf 87C} (1982) 77;\\ {\rm
  S.~Narison}, Phys.~Lett.~{\bf B216} (1989) 191;\\ {\rm N.~Gray,
  D.J.~Broadhurst, W.~Grafe and K.~Schilcher}, Z.~Phys.~{\bf C48} (1990) 673.}

\bibitem{hemp}
R.~Hempfling.
\newblock {\em Phys. Rev. {\bf D49} (1994) 6168}.

\bibitem{hall}
U.~Sarid L.~Hall, R.~Rattazzi.
\newblock {\em LBL-33997,UCB-PTH-93/14, Phys. Rev. {\bf D50}(1994)7048}.

\bibitem{ckw93}
K.G. Chetyrkin and A.~Kwiatkowski.
\newblock {\em Phys.~Lett.~{\bf B305} (1993) 285.}

\bibitem{higgslim}
{D.~Buskulic et al., ALEPH Coll.}
\newblock {\em Phys.~Lett.~{\bf B313} (1993) 312.}

\bibitem{sopczak}
{A.~Sopczak}.
\newblock {\em CERN-PPE/94-73,Lisbon Fall School 1993}.

\bibitem{borz}
{\em {\em F. Borzumati,} Z. Phys. {\bf C63} (1995) 291.}

\bibitem{bsgamm3}
{\em {\em S. Bertolini, F. Borzumati, A.Masiero, and G. Ridolfi}, Nucl. Phys.
  {\bf B353} (1991) 591 {\em and references therein;}\\ {\em N. Oshimo}, Nucl.
  Phys. {\bf B404} (1993) 20.}

\bibitem{burasb}
A.J.~Buras et~al.
\newblock {\em Nucl. Phys. {\bf B424}(1994)374}.

\bibitem{bsgamma}
{\em {\em R. Barbieri and G. Giudice,} Phys. Lett. {\bf B309} (1993) 86;\\ {\em
  R. Garisto and J.N. Ng,} Phys. Lett. {\bf B315} (1993) 372.}

\bibitem{alibsg}
A.~Ali and C.~Greub.
\newblock {\em Z. Phys. {\bf C60} (1993) 433}.

\bibitem{bop}
{\em {\em F.M.~Borzumati, M. Olechowski and S. Pokorski,} Phys. Lett. {\bf
  B349} (1995) 311.}

\bibitem{bsgamm1}
{\em {\em J. L. Lopez, D.V. Nanopoulos, G. T. Park,} Phys. Rev. {\bf D48}
  (1993) 974; {\em J.L. Lopez et al.,} Phys. Rev. {\bf D51} (1995) 147.}

\bibitem{bsgamm2}
{\em {\em J.L. Hewett}, Phys. Rev. Lett. {\bf 70} (1993) 1045;\\ {\em V.
  Barger, M.S. Berger, and R.J.N. Phillips,} Phys. Rev. Lett. {\bf 70} (1993)
  1368;\\ {\em M.A. Diaz}, Phys. Lett. {\bf B304} (1993) 278.}

\bibitem{borner}
G.~B\"orner.
\newblock {\em The early Universe,}.
\newblock Springer Verlag, (1991).

\bibitem{kolb}
E.W. Kolb and M.S. Turner.
\newblock {\em The early Universe,}.
\newblock Addison-Wesley, (1990).

\bibitem{relic}
{\em {\em G. Steigman, K.A. Olive, D.N. Schramm, M.S. Turner,} Phys. Lett. {\bf
  B176} (1986) 33; \\ {\em J. Ellis, K. Enquist, D.V. Nanopoulos, S. Sarkar, }
  Phys. Lett. {\bf B167} (1986) 457;\\ {\em G. Gelmini and P. Gondolo,} Nucl.
  Phys. {\bf 360} (1991) 145.}

\bibitem{relictst}
{\em {\em M. Drees and M. M. Nojiri,} Phys. Rev. {\bf D47} (1993) 376;\\ {\em
  J. L. Lopez, D.V. Nanopoulos, and H. Pois,} Phys. Rev. {\bf D47} (1993)
  2468;\\ {\em P. Nath and R. Arnowitt,} Phys. Rev. Lett. {\bf 70} (1993)
  3696;\\ {\em J. L. Lopez, D.V. Nanopoulos, and K. Yuan,} Phys. Rev. {\bf D48}
  (1993) 2766.}

\bibitem{rosdm}
{\em {\em L. Roszkowski, Univ. of Michigan Preprint}, UM-TH-93-06;
  UM-TH-94-02.}

\bibitem{lspdark}
{\em {\em J. Ellis et al.}, Nucl. Phys. {\bf B238} (1984) 453.}

\bibitem{susygen}
S.~Katsanevas.
\newblock {\em SUSYGEN,private communication}.

\bibitem{isajetee}
Baer et. al.
\newblock {\em Int.~Journ.~of~mod.~phys.Vol.4,16(1989)4111}.

\bibitem{isajet}
F.E Paige and S.D. Protopopescu.
\newblock {\em ISAJET7.11,Fermilab}.

\end{thebibliography}
%\input{biblio}

\end{document}